\documentclass[twocolumn,aps,prx,amsmath,amssymb,superscriptaddress,nofootinbib]{revtex4-1}
\usepackage{mathtools}
\usepackage{mathrsfs, amsmath, wasysym}
\usepackage[mathscr]{eucal}
\usepackage{xfrac}
\usepackage{graphicx}
\usepackage{dcolumn}
\usepackage{bm}
\usepackage{color}
\usepackage{tabularx}
\usepackage{natbib}
\usepackage[colorlinks,linkcolor=blue,citecolor=blue,urlcolor=blue]{hyperref}
\usepackage[normalem]{ulem}
\usepackage[all]{hypcap}
\usepackage[dvipsnames,usenames,table]{xcolor}

\newcommand{\nocontentsline}[3]{}
\newcommand{\tocless}[2]{\bgroup\let\addcontentsline=\nocontentsline#1{#2}\egroup}

\newcommand{\bk}{{\bf k}}

\newcommand{\bq}{{\bf q}}

\newcommand{\ba}{{\bf a}}

\newcommand{\bd}{{\bf d}}

\newcommand{\br}{{\bf r}}
\newcommand{\bR}{{\bf R}}

\newcommand{\bS}{{\bf S}}

\newcommand{\ephi}{\hat {\bf e}_\varphi}

\newcommand{\beq}{\begin{eqnarray}}
\newcommand{\eeq}{\end{eqnarray}}
\newcommand{\bea}{\begin{align}}
\newcommand{\eea}{\end{align}}
\newcommand{\beqq}{\begin{eqnarray*}}
\newcommand{\eeqq}{\end{eqnarray*}}

\newcommand{\makebf}[1]{\boldsymbol{#1} }

\newcommand{\bracket}[2]{\langle #1 | #2 \rangle }

\newcommand{\ket}[1]{ | #1 \rangle }

\newcommand{\up}{\uparrow}
\newcommand{\down}{\downarrow}

\makeatother

\begin{document}

\title{Correlations and electronic order in a two-orbital honeycomb lattice model \\
 for twisted bilayer graphene}

\author{J\"orn W. F. Venderbos}

\affiliation{Department of Physics and Astronomy, University of Pennsylvania,
Philadelphia, Pennsylvania 19104, USA}

\affiliation{Department of Chemistry, University of Pennsylvania, Philadelphia,
Pennsylvania 19104, USA}

\author{Rafael M. Fernandes}

\affiliation{School of Physics and Astronomy, University of Minnesota, Minneapolis,
Minnesota 55455, USA }

\date{\today}
\begin{abstract}
The recent observation of superconductivity in proximity to an insulating phase in twisted bilayer graphene (TBG) at small ``magic'' twist angles has been linked to the existence of nearly-flat bands, which make TBG a fresh playground to investigate the interplay between correlations and superconductivity. The low-energy narrow bands were shown to be well-described by an effective tight-binding model on the honeycomb lattice (the dual of the triangular Moir\'e superlattice) with a local orbital degree of freedom. In this paper, we perform a strong-coupling analysis of the proposed $\left(p_{x},\,p_{y}\right)$ two-orbital extended Hubbard model on the honeycomb lattice. By decomposing the interacting terms in the particle-particle and particle-hole channels, we classify the different possible superconducting, magnetic, and charge instabilities of the system. 
In the pairing case, we pay particular attention to the two-component ($d$-wave) pairing channels, which admit vestigial phases with
nematic or chiral orders, and study their phenomenology. Furthermore, we explore the strong-regime by obtaining a simplified spin-orbital exchange model which may describe a putative Mott-like insulating state at quarter-filling. Our mean-field solution reveals a rich intertwinement between ferro- and antiferro-magnetic orders with different types of nematic and magnetic orbital orders. Overall, our work provides a solid framework for further investigations of the phase diagram of the two-orbital extended Hubbard model in both strong- and weak-coupling regimes.
\end{abstract}
\maketitle

\section{Introduction}

The experimental discovery of superconductivity in twisted bilayer
graphene (TBG) \cite{cao2018a,cao2018b,mele2018} has attracted much
attention and has triggered a considerable theoretical effort to address
this unexpected observation \cite{xubalents2018,yuan2018,po2018a,roy2018,guo2018,baskaran2018,padhi2018,irkhin2018,dodaro2018,huang2018,zhang2018,ray2018,liu2018,xu2018,kang2018,rademaker2018,isobe2018,koshino2018,wumartin2018,pizarro2018,peltonen2018,you2018,wu2018,pal2018,ochi2018,fidrysiak2018,kennes2018,thomson2018,guinea2018,patel2018,zou2018,gonzalez2018,su2018,lian2018,sherkunov2018,sboychakov2018,song2018,chittari2018,hejazi2018,po2018b,tomanek2018,laksono2018,tarnopolsky2018,lin2018,ahn2018}.
In particular, a renewed interest in the low-energy electronic properties
of TBG structures has surfaced, geared towards incorporating correlations
on the electronic structure via controlled approaches. From a more
general perspective, the discovery of superconductivity on TBG has
brought back into focus long-standing and much-debated questions concerning
the interplay of electronic correlations and superconductivity \cite{carbotte1990,Lee_Nagaosa_Wen,scalapino2012,chubukov2013}.

\begin{figure}
\includegraphics[width=0.9\columnwidth]{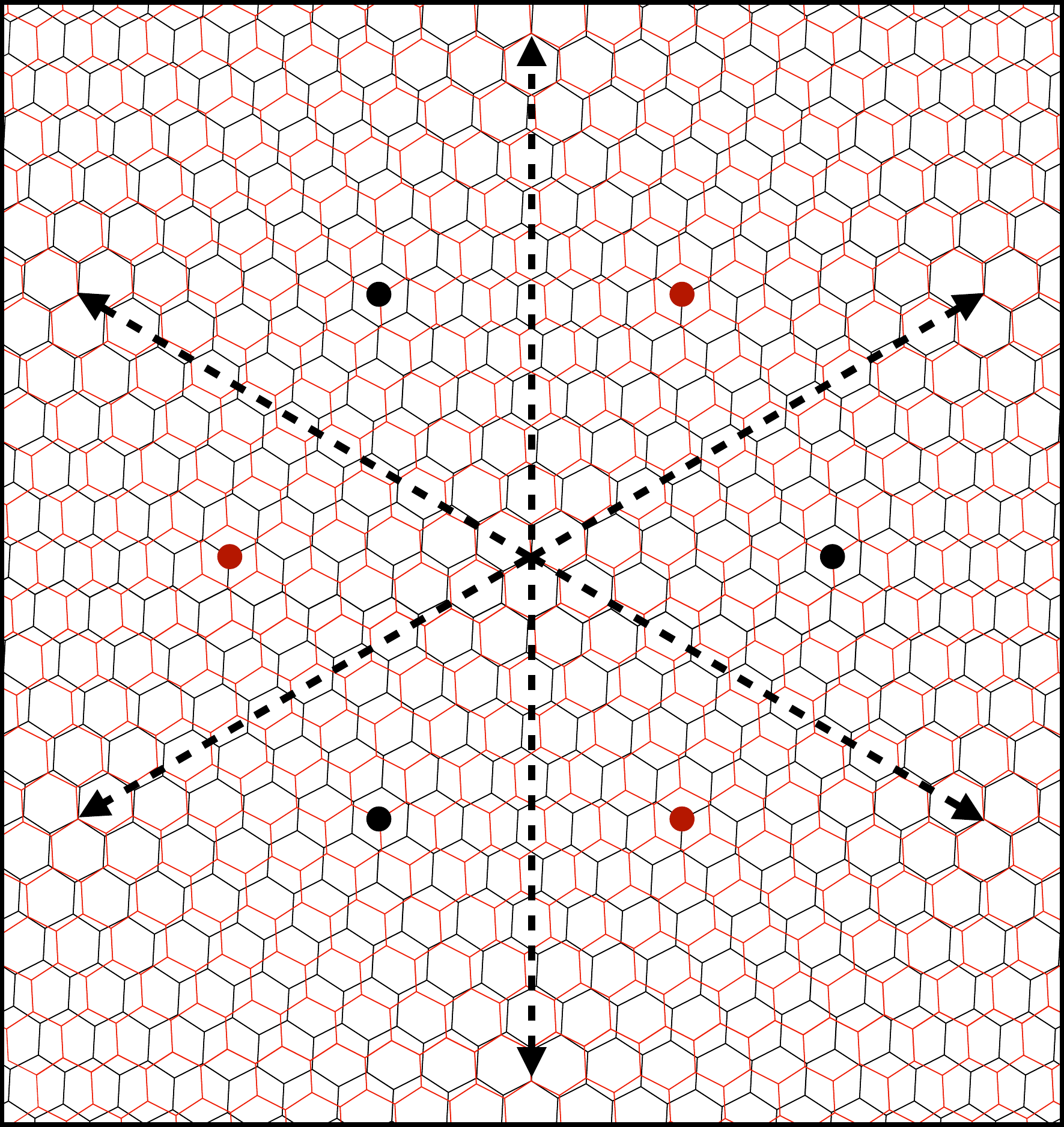} \caption{\label{fig:twisted} \textbf{Twisted bilayer graphene.} Figure of
two twisted graphene sheets, shown as black and red honeycomb nets,
with commensurate Moir\'e superlattice periodicity. In this commensurate
realization of twisted bilayer graphene, the twist angle is $\theta=6.01^{\circ}$
and the twist center is a pair of registered carbon atoms which defines
the origin. The triangular superlattice vectors connecting regions
of $AA$ stacking are shown by dashed arrows. The black and red dots
indicate the sites of the dual honeycomb (super)lattice and correspond
to regions of $AB$ and $BA$ stacking, respectively. }
\end{figure}

Twisted bilayer graphene belongs to the class multilayer graphene
systems generated by stacking sheets of monolayers. Given the large
set of distinct stacking prescriptions, multilayer graphene systems
offer a high degree of tunability of the resulting electronic structure
\cite{nilsson2006,nilsson2008}. By stacking two graphene sheets to
form a bilayer and rotating (``twisting'') one layer with respect
to the other by an angle $\theta$, one obtains a triangular Moir\'e
superlattice structure (shown in Fig.~\ref{fig:twisted}) \cite{mccann2013,rozhkov2016}.

Based on an experimental study of TBG with small twist angles $\theta\sim1^{\circ}$,
Cao \emph{et al.} reported a metal-to-insulator transition at $T\approx4$
K for carrier densities corresponding to $\pm2e$ per Moir\'e supercell
(with respect to charge neutrality) \cite{cao2018a}. The conductance
in the insulating state displays activated behavior with an activation
energy $\Delta\approx0.3$ meV, comparable to the metal-to-insulator
transition temperature. Remarkably, upon doping slightly away from
$\pm2e$ per supercell, either by adding holes or electrons, a superconducting
state with a maximum transition temperature of $T_{c}\approx1.7$
K was observed \cite{cao2018b}. In fact, even the half-filled system
was found to superconduct at low temperatures in the absence of a
magnetic field for certain values of $\theta$. The existence of superconductivity
near an insulating state was also reported in Ref. \onlinecite{Dean2018},
where pressure was used to tune the ground state of TBG with larger
twist angles.

These observations raise important questions about the nature of the
insulating and superconducting states, as well as the interplay between
them. The fact that the insulating state appears at densities where
single-particle considerations would predict metallic behavior, hints
at the importance of electronic correlations. Indeed, for twist angles
$\theta\sim1^{\circ}$ numerical calculations had previously predicted
the existence of Moir\'e minibands with almost flat dispersion near
the Fermi level \cite{shallcross2010,morell2010,trambly2010,jung2014}.
Some works reported a set of four narrow-bandwidth minibands (eight
including spin degeneracy) separated from the other bands above and
below \cite{fang2016,namkoshino2017}, which appears to agree with
the experimental findings. The small bandwidth $W\sim10$ meV of this
set of low-energy bands suggests that correlations are likely to provide
the dominant energy scale and drive the system into a Mott-like state
at quarter filling.

On the other hand, the fact that the insulating transport behavior
only onsets at relatively low-temperatures comparable to $T_{c}$,
combined with the small magnetic fields needed to kill the insulating
state (of the order of $4$ Tesla), can be viewed as a challenge to
the Mott-like scenario \cite{padhi2018,pizarro2018,dodaro2018}. As
a result, alternative explanations for the insulating state have been
put forward \cite{dodaro2018,liu2018,xu2018,isobe2018,thomson2018,ochi2018,sboychakov2018}. Regardless of the microscopic origin of the insulating state, the onset of a relatively high $T_{c}$ state
at its vicinity and at such low densities hint at the possibility
of unconventional electronically-driven pairing.

\begin{figure}
\includegraphics[width=\columnwidth]{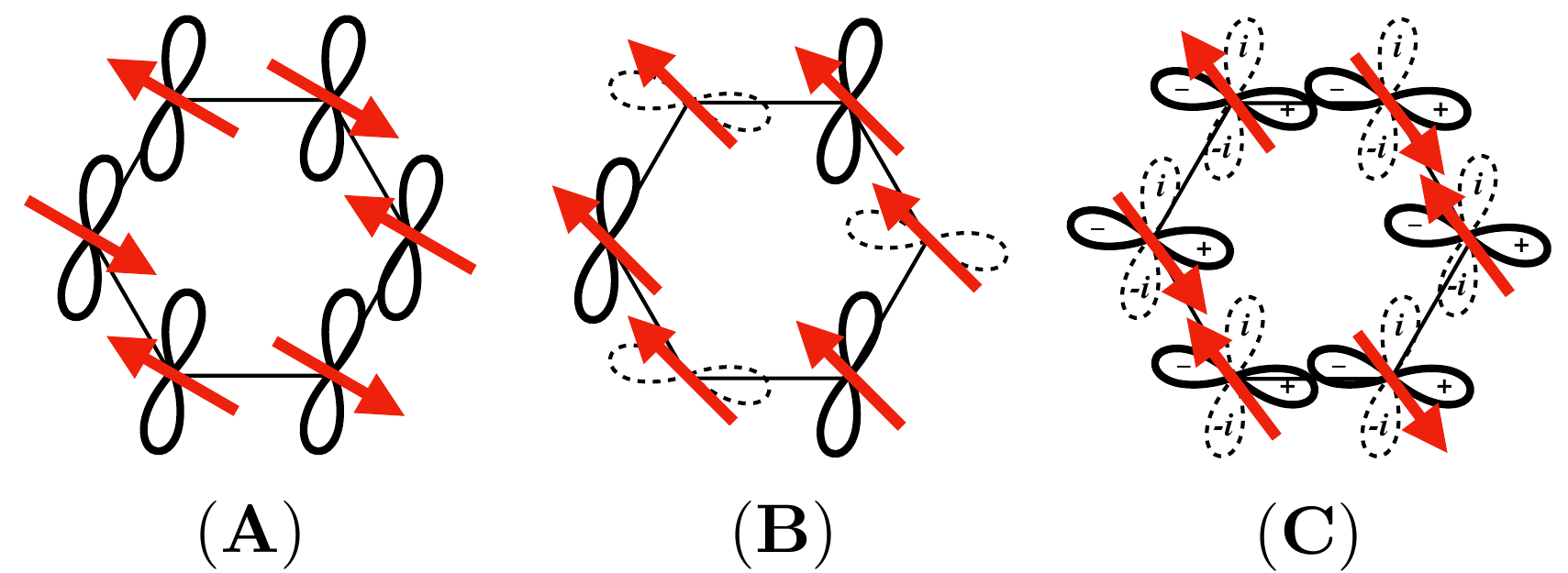} \caption{\label{fig:orderings} \textbf{Orbital and spin ordering.}  Schematic picture of the intertwined spin and orbital orderings appearing in the Mott insulating state at quarter-filling, as discussed in Sec.~\ref{sec:spin-orbital}. Solid and dashed orbitals refer to the different $p_y$ and $p_x$ orbitals. ($\bf A$) Antiferromagnetic ferro-orbital order; ($\bf B$) Ferromagnetic antiferro-orbital order; ($\bf C$) Antiferromagnetic ferro-orbital-magnetic order with complex orbitals.}
\end{figure}

To answer these questions, appropriate models to describe the electronic
structure are needed. Studies of TBG structures predating the recent
experimental reports have addressed the electronic properties of TBG
primarily within the framework of a low-energy continuum model, which
starts from the Dirac electrons of the individual graphene layers
\cite{lopes2007,mele2010,mele2011,bistritzer2011,lopes2012,koshino2012}.
This has proven to provide an excellent description for the low-energy
electronic structure, in particular the appearance of nearly flat
bands at charge neutrality, manifested by a vanishing of the Fermi
velocity at special (``magic'') twist angles. Since the manifold
of nearly-flat low-energy bands at charge neutrality is well-separated
from other bands, a description which accurately captures these bands
may be sufficient. 

Therefore, more recent works \cite{yuan2018,po2018a,zhang2018,kang2018,rademaker2018,koshino2018,zou2018,song2018,po2018b,tomanek2018,ahn2018}
have set out to formulate an effective tight-binding lattice model
akin to (multi-orbital) Hubbard models. The construction of an effective tight-binding model for the
nearly-flat bands, which relies on extracting localized Wannier states
from the miniband structure, was shown to be contingent on the (exact
and approximate) symmetries that are imposed on the model~\footnote{A brief discussion of the intricacies involved in the Wannier state
construction is given un Sec.~\ref{sec:model}, with directions to
the relevant references.}. What is perhaps most important, however, is that any consistent
formulation of a tight-binding model in terms of Wannier states was
shown to require a honeycomb lattice structure \cite{yuan2018,po2018a}. Whereas the
triangular Moir\'e lattice can be defined by regions of $AA$ stacking,
the dual honeycomb lattice is defined by regions of $AB$ and $BA$
stacking (see Fig. \ref{fig:twisted}).

In this paper, we start from the extended two-orbital Hubbard model
proposed in Refs. \cite{yuan2018,kang2018,koshino2018} and explore
the effect of correlations on the low-energy flat bands. In this model,
the orbitals have $(p_{x},p_{y})$ symmetry and one of 
our main goals is to assess the role of the $(p_{x},p_{y})$
orbital degrees of freedom on the superconducting, charge, and magnetic
instabilities of the model. 
Here,
we first decompose the interacting part of the Hubbard model, which
involves both onsite and longer-range interactions, in the particle-particle
and particle-hole channels. In this way, we obtain a general symmetry classification of pairing and particle-hole instabilities, which allows us to determine the effective interaction in each irreducible channel. The latter reveals which channels are most attractive (or least repulsive). 
In the case of pairing, we pay particular attention to the two-component ($d$-wave) superconductivity, which supports vestigial
non-superconducting states with either chiral or nematic order. We argue that TBG is an ideal candidate to
realize such vestigial states, given the reduced dimensionality of
the system.

Having decomposed the interactions into irreducible channels, one
can include the contributions from the kinetic term by either treating
the kinetic part perturbatively (strong-coupling) or the interaction
terms perturbatively (weak-coupling). In this paper, motivated by
the small bandwidth of the low-energy flat bands, we explore the former
regime, but we emphasize that the same formalism can also be used
for weak-coupling analyses. Here we focus on the putative Mott state at quarter-filling and consider an (anisotropic) spin-orbital
exchange model, analogous to the Kugel-Khomskii-type Hamiltonians~\cite{kugel1973,kugel1982}
commonly employed to describe strongly correlated multi-orbital systems~\cite{imada1998,dagotto2001,dagotto-book,meakawa-book}.
As a first step towards
understanding the implications of such spin-orbital Hamiltonian, we
perform a mean-field analysis in the case where only onsite interaction
terms are kept. Depending of the ratio between the Hund's coupling
$J$ and the Hubbard $U$, we find antiferromagnetic order coupled
either to a ferro-orbital nematic order or to a ferro-orbital magnetic
order, or ferromagnetic order coupled to an $SU(2)$ antiferro-orbital
order. A schematic representation of these results is shown in Fig.~\ref{fig:orderings}.

As mentioned before, the Mott scenario should and will be subject
to critical discourse. Insofar as the derivation and analysis of a
spin-orbital exchange Hamiltonian is concerned, two important qualifying
remarks are worth making. First, we note that in the derivation of
such Hamiltonian only onsite repulsion is considered. In the context
of TBG this is a rather restrictive assumption, since the structure
of the orbital Wannier states suggests that farther neighbor repulsion
is non-negligible~\cite{koshino2018,ochi2018,pizarro2018}. Second,
the assumption of a small bandwidth $W$ as compared to the (onsite)
interaction energy scale $U$, i.e., $W/U\ll1$, seems questionable
given the small value of the activation transport gap $\Delta$ and
the low temperature at which the metal-to-insulator transition takes
place. Nevertheless, a careful examination of strong-coupling approaches
to TBG are expected to offer interesting and important insight into
the correlated physics of TBG.

The paper is organized as follows: Sec.~\ref{sec:model} introduces
and discusses the extended two-orbital Hubbard model with an emphasis
on its symmetries. This section is largely a review of the studies
which have proposed and constructed the two-orbital honeycomb lattice
model, but we believe a thorough discussion may benefit the reader.
In Secs.~\ref{sec:sc} and~\ref{sec:p-h-order} the pairing instabilities
and particle-hole instabilities are considered, respectively, by decomposing
the interacting part of the Hamiltonian into irreducible superconducting
and particle-hole channels. In Sec. \ref{sec:spin-orbital}, the kinetic
part is included perturbatively, and the resulting spin-orbital exchange
model is derived and analyzed within a mean-field approach. Sec. \ref{sec:conclusion}
is devoted to concluding remarks. A number of Appendices, Appendix
\ref{app:wannier}\textendash \ref{app:tau}, collect additional details
of the calculations presented in the main text.

\section{Low-energy two-orbital Hubbard model for twisted bilayer graphene
\label{sec:model}}

\subsection{General considerations \label{ssec:symmetries}}

Our starting point is the effective extended Hubbard model for the
low-energy flat bands of TBG developed in a series of recent works~\cite{yuan2018,po2018a,kang2018,koshino2018,zou2018}.
The effective tight-binding model for the flat-band manifold takes
the form of a honeycomb lattice model with two Wannier orbitals per
honeycomb lattice site, which was demonstrated based on a symmetry
analysis~\cite{yuan2018} and an explicit calculation of maximally
localized Wannier orbital wavefunctions~\cite{kang2018,koshino2018}.
The Bravais lattice vectors of the honeycomb lattice correspond to
the lattice vectors of the triangular Moir\'e superlattice generated
by the twist. The sites of the triangular Moir\'e superlattice can be
identified with regions of local $AA$ stacking, whereas the sublattices
of the honeycomb lattice, which is the dual of the triangular lattice,
mark the centers of local $AB$ and $BA$ stacking, respectively.
This is shown in Fig.~\ref{fig:twisted}. Note that the structure
of the honeycomb lattice implies four orbitals in the superlattice
unit cell, i.e., two Wannier states per sublattice, which is consistent
with the number of nearly-flat bands forming the low-energy manifold.
Importantly, in such a superlattice model the two Wannier orbitals
transform in a specific way under spatial symmetries of TBG and these
symmetry properties dictate the form of the hopping and interaction
terms of the effective tight-binding model. For instance, in some
cases the Wannier states were shown to transform as $p$-wave partners
under rotations~\cite{yuan2018,kang2018,koshino2018}.

The construction of the honeycomb superlattice tight-binding model,
and in particular the derivation of the localized Wannier functions,
is predicated on two important assumptions, which are useful to state
explicitly. The first assumption is the existence of exact lattice
translation and point group symmetries of TBG. The presence of exact
translational symmetry of the twisted structure implies a commensurability
condition on the Moir\'e supercell, which in turn implies a constraint
on the twist angle $\theta$. 
Note that for small but commensurate twist angles the unit cell of the Moir\'e superlattice
unit cell can become very large.

\begin{figure}
\includegraphics[width=1\columnwidth]{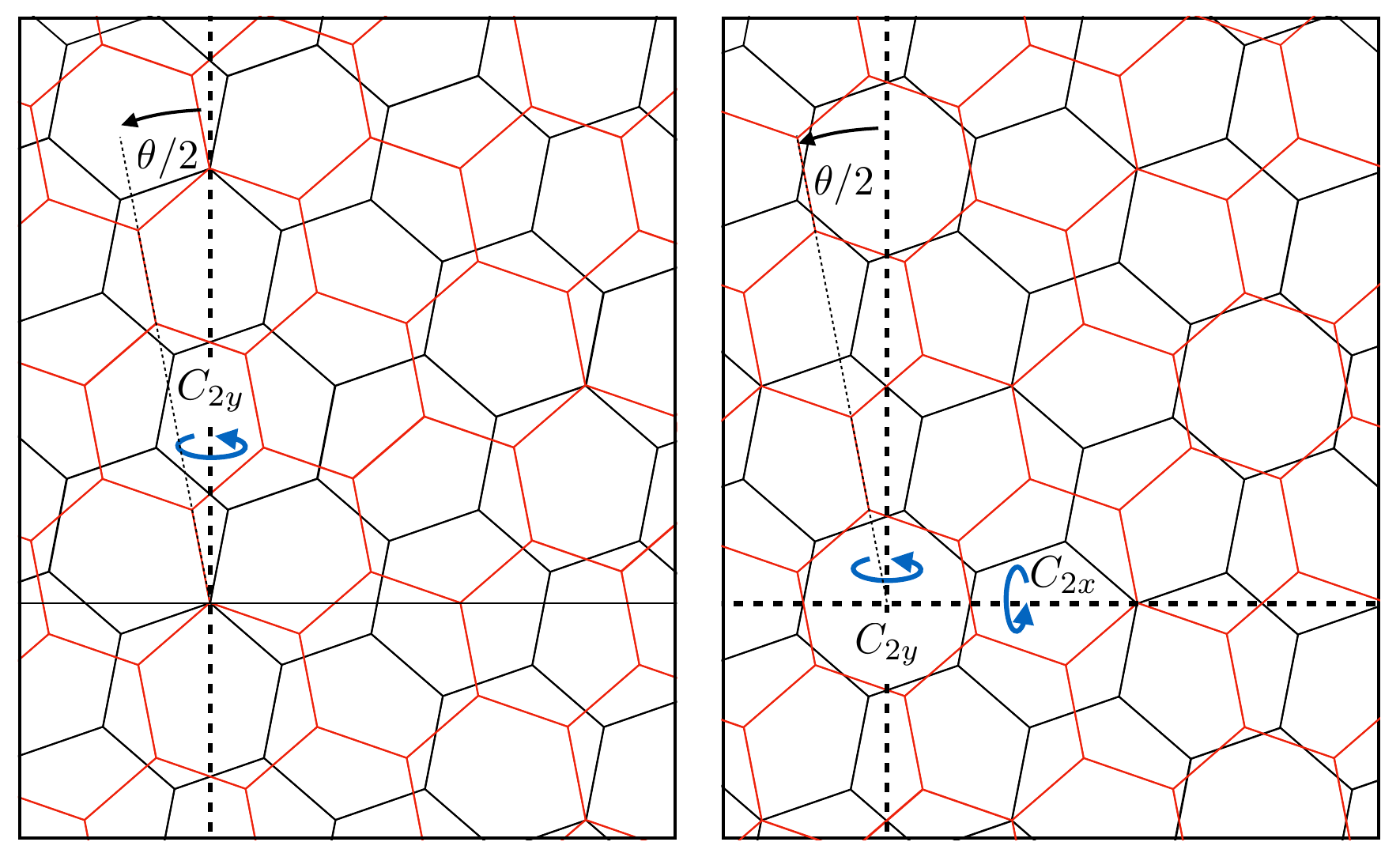} \caption{\label{fig:lattice} \textbf{Symmetry of TBG.} (Left panel) Example
of TBG structure with $D_{3}$ point group symmetry. The twist rotation
axis is coincident with a pair of registered carbon atoms. The structure
has a two-fold rotational symmetry $C_{2y}$ about the $y$ axis,
and $C_{3z}$ three-fold rotational symmetry about the $z$ axis.
(Right) For comparison, we show a TBG structure where the twist rotation
axis is coincident with the center of a hexagon, resulting in $D_{6}$
point group symmetry. This implies an additional two-fold rotational
symmetry $C_{2x}$, and a $C_{6z}$ six-fold rotational symmetry about
the $z$ axis. Both structures, left and right, have the same twist
angle (and Moir\'e period), which was chosen large for illustrative
purposes. Importantly, the twist center is also the center for the
$C_{3z}$ rotations, both for the $D_{3}$ and $D_{6}$ structures. }
\end{figure}

In addition to translational symmetry, the construction of the tight-binding
model also assumes the existence of an exact point group symmetry.
Indeed, the aforementioned statement that the Wannier orbitals (in
some cases) have $p$-wave symmetry can only have meaning when rotational
symmetry is present. Commensurate TBG structures can belong to one
of two possible dihedral point groups: $D_{3}$ or $D_{6}$. The difference
in rotational symmetry depends on the center of twist rotation, as
illustrated in Fig. \ref{fig:lattice}. To understand this difference,
consider starting from two $AA$ stacked graphene sheets and rotating
the top (bottom) layer by an angle $\theta/2$ ($-\theta/2$) about
an axis coincident with two registered carbon atoms, with $\theta$
defined with respect to the $y$ axis. This results in a structure
with three-fold rotation symmetry $C_{3z}$ along the $z$ axis and
two-fold rotation symmetry $C_{2y}$ along the $y$ axis, as shown
in the left panel of Fig.~\ref{fig:lattice}. Together these two
symmetries generate $D_{3}$. The TBG structure shown in Fig. \ref{fig:twisted}
is an example of the latter. Alternatively, if the twist rotation
axis is coincident with the center of graphene hexagons, shown in
the right panel of Fig.~\ref{fig:lattice}, the resulting TBG structure
retains the six-fold $C_{6z}$ rotation symmetry; in combination with
$C_{2y}$ this generates $D_{6}$ (which includes the twofold rotation
$C_{2x}$). 

A second important assumption of the Wannier orbital construction
is the existence of an energy gap between the four flat bands and
the other bands. 
The existence of such an energy gap has been
predicted by theory \cite{fang2016,namkoshino2017} and appears to be consistent with experiment \cite{cao2018a}.

Following these considerations, we now introduce the honeycomb (super)lattice
tight-binding model on which our study is based. The honeycomb lattice model we focus on in this work is meant to describe commensurate TBG structures
with $D_{3}$ symmetry, shown in Fig. \ref{fig:lattice} on the left. It was shown that for this case the two Wannier orbitals at each superlattice
site transform as two $p$-wave states~\cite{yuan2018,kang2018,koshino2018}. This is a particularity of the $D_3$ symmetric structures, for which the center
of the $C_{3z}$ rotation is defined by registered carbon atoms~\cite{po2018a,zou2018}. In the case of commensurate structures
with $D_{6}$ symmetry, the symmetry quantum numbers of the Wannier states were found to be different~\cite{po2018a}, resulting in a
different tight-binding description of the low-energy flat bands. More generally, the construction of Wannier states depends on the exact and approximate symmetries of TBG which are imposed on the construction. Unless some (approximate) symmetries are ignored, the construction of localized symmetric Wannier states is obstructed~\cite{po2018a,zou2018}. Here we do not give a full account of the subtleties and caveats
related to construction of Wannier orbitals, in particular to the
(exact or emergent) symmetries which are imposed, but instead refer
the reader to the relevant Refs. \onlinecite{yuan2018,po2018a,kang2018,koshino2018,zou2018,song2018}, in particular Ref.~\onlinecite{po2018a}.

\subsection{Two-orbital extended Hubbard model \label{ssec:model}}

Given the symmetry of the Wannier states we denote the orbitals at
each site $i$ as $p_{x,y}$ and define the corresponding electron
annihilation (creation) operators as $c_{i\alpha\sigma}$ ($c_{i\alpha\sigma}^{\dagger}$)
with $\alpha=x,y$ and $\sigma=\up,\down$ for spin. The kinetic part
of the Hamiltonian describes the hopping processes and can be expressed
as 
\begin{equation}
H_{K}=\sum_{ij}c_{i}^{\dagger}\hat{T}(\br_{ij})c_{j}+\text{h.c.},\label{eq:H-K}
\end{equation}
where $\hat{T}(\br_{ij})$ are hopping matrices and $\br_{ij}=\br_{i}-\br_{j}$
is the distance between sites $i$ and $j$. Spin-orbit coupling is
neglected, giving rise to full $SU(2)$ spin rotational invariance.

\begin{figure}
\includegraphics[width=1\columnwidth]{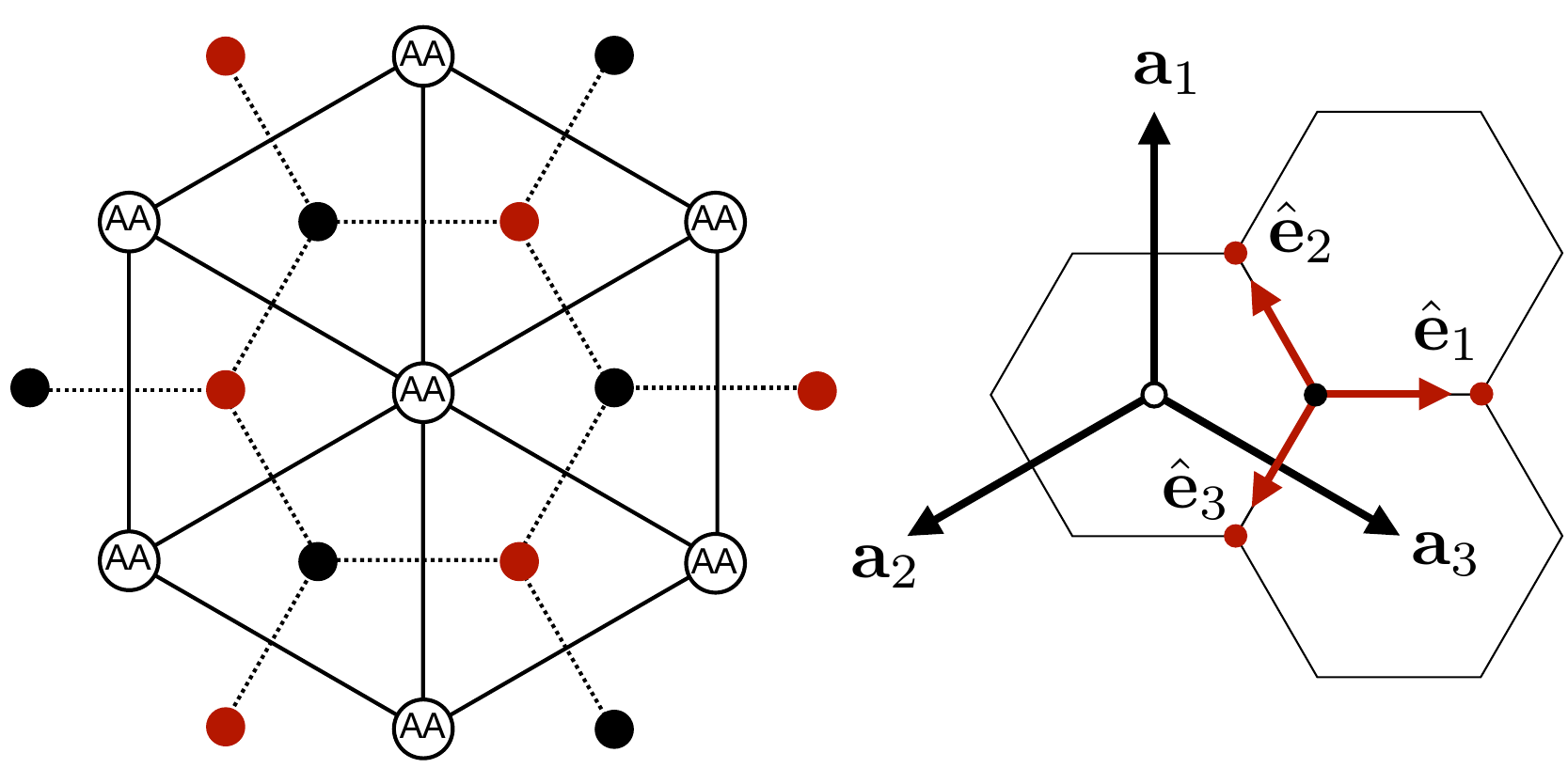} \caption{\label{fig:lattice_bonds} \textbf{Honeycomb superlattice model.}
(Left) Sketch of the effective honeycomb lattice extracted from twisted
bilayer graphene with commensurate twist angle (see Fig.~\ref{fig:twisted}).
The triangular Moir\'e superlattice, defined by the regions of $AA$
stacking, is shown by solid lines. Red and black solid dots represent
the sites of the honeycomb lattice (indicated by dashed lines), with
different colors corresponding to the triangular sublattices of the
honeycomb lattice. The sublattice sites coincide with regions of $AB$
and $BA$ stacking. (Right) Definition of lattice vectors. Here $\ba_{1,2,3}$
are lattice vectors of the (triangular) Moir\'e superlattice and $\hat{{\bf e}}_{1,2,3}$
are unit vectors corresponding to the directions of nearest-neighbor
bonds.  
}
\end{figure}

For each set of bonds with fixed $\br_{ij}$ (i.e. nearest neighbors,
next-nearest neighbors, etc.) the form of the hopping matrices is
constrained by the transformation properties of the $p_{x,y}$ orbitals
states under the $D_{3}$ point group symmetry. Time-reversal symmetry
imposes an additional constraint on the hopping matrices. A 
derivation of the symmetry constraints on the hopping matrices was presented in Ref.~\onlinecite{kang2018}; here, we review this briefly using a different formalism, with details given in Appendix~\ref{app:symmetryD3}. To 
exploit rotational symmetry, we introduce a set of unit vectors corresponding to the bond directions; first, we define a general rotated frame 
\begin{equation}
\hat{{\bf e}}_{\varphi}=\cos\varphi\hat{{\bf e}}_{x}+\sin\varphi\hat{{\bf e}}_{y},\quad\hat{{\bf e}}_{\varphi}^{\perp}=-\sin\varphi\hat{{\bf e}}_{x}+\cos\varphi\hat{{\bf e}}_{y},\label{eq:e-theta}
\end{equation}
where $\varphi$ is an arbitrary angle and $\ephi\times\ephi^{\perp}=\hat{{\bf e}}_{z}$.
The three nearest neighbor unit vectors are then specified by $\varphi_{n}=2\pi(n-1)/3$.
We define the nearest neighbor unit vectors as $\hat{{\bf e}}_{n=1,2,3}$,
see Fig.~\ref{fig:lattice_bonds}, and denote the corresponding hopping
matrices as $\hat{T}_{n=1,2,3}^{(1)}$. Since the three hopping matrices
are related by threefold rotations only one needs to be specified.
Focusing on $\hat{T}_{1}^{(1)}$, we find: 
\begin{equation}
\hat{T}_{1}^{(1)}=t_{1}+t'_{1}\tau^{z}.\label{eq:T-1}
\end{equation}
Here the Pauli matrices $\tau^{x,y,z}$ act on the orbital degrees
of freedom, i.e., $\tau^{z}=\pm1$ corresponds to $p_{x,y}$. Note
that the hopping matrix along the nearest neighbor bond direction
$\hat{{\bf e}}_{n=1}$ is diagonal in orbital space. By analogy with
atomic $p$-orbitals, we may introduce $\sigma$- and $\pi$-hopping
processes as $t_{\sigma,\pi}=t_{1}\pm t'_{1}$. The computation of
$\hat{T}_{n=2,3}^{(1)}$ follows from \eqref{eq:T-1} by appropriate
rotations, as outlined in Appendices \ref{app:wannier} and \ref{app:symmetryD3}.

Importantly, to reproduce details of the band structure of TBG longer
ranged hopping processes must be included~\cite{yuan2018,kang2018,koshino2018}, in particular 
intra-sublattice hopping matrices, i.e., hopping matrices connecting two sites on the
same triangular sublattice. The most important hopping processes of
this kind are second-nearest and fifth-nearest neighbor hopping. Viewed
as bonds on the triangular sublattice these are first-nearest and
second-nearest neighbor hoppings. We introduce the hopping matrices
$\hat{T}_{n=1,2,3}^{(2)}$ and $\hat{T}_{n=1,2,3}^{(5)}$, with $\hat{T}_{1}^{(2)}$
in the direction of $\hat{{\bf e}}_{y}$ and $\hat{T}_{1}^{(5)}$
in the direction of $\hat{{\bf e}}_{x}$. (Note that the three second-nearest
neighbor bonds correspond to $\ba_{1,2,3}$, as shown in Fig.~\ref{fig:lattice_bonds}.)
Examining the constraints from symmetry, we arrive at (see Appendix~\ref{app:symmetryD3})
\begin{eqnarray}
\hat{T}_{1}^{(2)} & = & t_{2}+t_{2z}\tau^{z}\pm t_{2x}\tau^{x}\pm it_{2y}\tau^{y},\label{eq:T-2}\\
\hat{T}_{1}^{(5)} & = & t_{5}+t_{5z}\tau^{z}\pm t_{5x}\tau^{x}+it_{5y}\tau^{y},\label{eq:T-5}
\end{eqnarray}
where $+$ ($-$) applies to the honeycomb sublattice $A$ ($B$).
As before, all other hopping matrices are obtained from rotation.

In principle, a symmetry analysis of this kind can be applied to any
hopping process of arbitrary range, resulting in the most general
form of Eq~\eqref{eq:H-K} consistent with symmetry.

The relation of this two-orbital honeycomb lattice model to the underlying degrees of
freedom of the individual graphene sheets (e.g., layer, sublattice,
valley) deserves further discussion. We mentioned that the two sublattices
of the Moir\'e honeycomb lattice, distinguished by black and red sites
in Fig. \ref{fig:lattice_bonds}, may be identified with regions of
$AB$ and $BA$ stacking of the graphene layers, where $A$ and $B$
refer to the sublattice degree of freedom of each graphene sheet.
There are thus two distinct notions of a sublattice degree of freedom,
which should not be confused. Unless otherwise specified, in what
follows the sublattice degree of freedom will be understood to refer
to the emergent honeycomb superlattice.

More importantly, even though the two Wannier states have $p$-wave
symmetry, which warrants the notation $p_{x,y}$, they should be clearly
distinguished from physical atomic $p_{x,y}$-orbitals. This is evidenced by the fact that the hopping
parameters of Eqs.~\eqref{eq:T-1}\textendash \eqref{eq:T-5}, in
particular the overlap integrals $t_{\sigma,\pi}=t_{1}\pm t'_{1}$
of Eq. \eqref{eq:T-1}, are not determined by the Slater-Koster rules~\cite{slater1954}.
Indeed, application of the Slater-Koster rules would imply Hermitian
hopping matrices. The hopping parameters can be directly
calculated from the Wannier states, which were shown to have maxima at located at the $AA$
stacking regions that form the triangular Moir\'e superlattice \cite{kang2018,koshino2018}.
In particular,
Ref. \onlinecite{koshino2018} demonstrated that: \textit{(i)} the
Wannier states have spectral weight on both layers and both sublattices
of each graphene layer; and \textit{(ii)} the Wannier states can be
associated with the valley degree of freedom of the constituent graphene
layers~\cite{note3}. This correspondence can be stated more precisely
by forming the complex Wannier orbitals $p_{\pm}=p_{x}\pm ip_{y}$
and noting that, within the approach followed by Ref.~\onlinecite{koshino2018},
$p_{+}$ and $p_{-}$ derive from valleys $K$ and $K'=-K$, respectively.
Furthermore, since the complex orbitals are eigenstates of $\tau^{y}$,
it is straightforward to see that if the hopping matrices of Eq. \eqref{eq:H-K}
{[}and in particular those of Eqs. \eqref{eq:T-1}\textendash \eqref{eq:T-5}{]}
only have nonzero terms proportional to the identity and $\tau^{y}$,
a larger internal $U(1)$ symmetry in orbital space emerges, generated
by $\tau^{y}$.

By calculating the overlap between Wannier orbitals, both Ref. \onlinecite{kang2018}
and \onlinecite{koshino2018} found that this larger $U(1)$ symmetry
is a good approximate symmetry of the tight-binding model, although
not exact. For Eq. \eqref{eq:T-1}, for instance, this implies $t_{\sigma}\approx t_{\pi}$
(i.e. $t'_{1}\ll t_{1}$). In addition, the importance of further
neighbor hopping terms was established, which can be traced back to
the real space extension of the Wannier states. We thus conclude that
TBG with exact $D_{3}$ symmetry is well-described by a kinetic tight-binding
Hamiltonian \eqref{eq:H-K} with longer ranged hoppings and an approximate
$U(1)$ symmetry, which can be associated with the valley quantum
number. We note in passing that for a rather different set of parameter,
i.e. only nearest neighbor $\sigma$-hopping ($t'_{1}=t_{1}$), the
physics of the honeycomb lattice $p$-orbital model was shown to give
rise to interesting physics, albeit most likely not relevant to TBG
\cite{wu2007,wu2008}.

Next, we consider the interacting part of the Hamiltonian, $H_{I}$.
The interacting Hamiltonian may be viewed as a sum of two types of
terms: density-density interaction terms and exchange terms. In its
most general form, $H_{I}$ is given by 
\begin{multline}
H_{I}=\frac{1}{2}\sum_{ij}V_{ij}^{\alpha\beta}n_{i\alpha}n_{j\beta}+\frac{1}{2}\sum_{ij,\alpha\beta}J_{1,ij}^{\alpha\beta}c_{i\alpha\sigma}^{\dagger}c_{j\beta\sigma'}^{\dagger}c_{i\beta\sigma'}c_{j\alpha\sigma}\\
+\frac{1}{2}\sum_{ij,\alpha\neq\beta}J_{2,ij}^{\alpha\beta}c_{i\alpha\sigma}^{\dagger}c_{j\beta\sigma'}^{\dagger}c_{i\alpha\sigma'}c_{j\beta\sigma}\\
+\frac{1}{2}\sum_{ij,\alpha\neq\beta}J_{3,ij}^{\alpha\beta}c_{i\alpha\sigma}^{\dagger}c_{j\alpha\sigma'}^{\dagger}c_{i\beta\sigma'}c_{j\beta\sigma},\label{eq:H-I}
\end{multline}
where the first term describes density-density interactions and the
remaining three terms describe exchange interactions. The four sets
of interaction parameters are not fully independent, but must satisfy
the constraint of invariance under rotations in orbital space (for
a formulation of this constraint see Appendix~\ref{app:decomposition}).
For each set of the interaction parameters we furthermore assume $V_{ij}^{xy}=V_{ij}^{yx}$
and $V_{ij}^{xx}=V_{ij}^{yy}$, and similarly for $J_{1,2,3}$. Finally,
we note that the interaction parameters are invariant under translations:
$V_{ij}^{\alpha\beta}\equiv V^{\alpha\beta}(\br_{i}-\br_{j})$, and
similarly for the exchange terms.

An extended Hubbard model of the form of Eq.~\eqref{eq:H-I} was
proposed in Ref. \onlinecite{koshino2018}, where the interaction
parameters were estimated using the Coulomb interaction and the explicit
wave-functions of the Wannier states. Such estimates showed that farther
neighbor interactions, while smaller than onsite interactions, are
non-negligible. In addition, in the context of the model used in Ref.
\onlinecite{koshino2018} the exchange interactions $J_{2,3}$ were
found to be considerably smaller than $J_{1}$. In Secs.~\ref{sec:sc}
and \ref{sec:p-h-order}, where we study the pairing and particle-hole
instabilities, we consider $H_{I}$ in its general form of Eq.~\eqref{eq:H-I}.
The main physical motivation to do so is that, because \eqref{eq:H-I}
is meant to describe the effective interactions within the manifold
of the low-energy flat bands, they are expected to get renormalized
by integrating out higher energy degrees of freedom (see, for instance,
Ref.~\onlinecite{chubukov2016}). 

In Sec.~\ref{sec:spin-orbital}, where we focus on the strong-coupling
regime, we study a particular limiting case of $H_{I}$ and only consider
the onsite interactions. Despite the fact that farther neighbor interactions
may not be too much smaller than the onsite terms, this approximation
is useful as it allows for the derivation of a spin-orbital exchange
Hamiltonian. Keeping onsite interactions only $(i=j)$ in Eq.~\eqref{eq:H-I},
the parameters $J_{1,ii}^{\alpha\beta}$ are equivalent to $V_{ii}^{\alpha\beta}$,
and the former may thus be set to zero. The remaining interaction
parameters can specified in terms of two interaction energy scales:
a Hubbard interaction $U$ and a Hund's rule coupling $J$ \cite{yuan2018}.
In terms of these two parameters, the non-zero onsite interaction
coefficients of Eq.~\eqref{eq:H-I} are $V^{xx}=V^{yy}=U$, $V^{xy}=V^{yx}=U-2J$,
and $J_{2,3}^{xy}=J_{2,3}^{yx}=J$. As a result, the Hamiltonian $H_{I}$
acquires the standard Hubbard-Kanamori form~\cite{kanamori1963}
\begin{multline}
H_{I}^{\mathrm{(onsite)}}=U\sum_{i,\alpha}n_{i\alpha\up}n_{i\alpha\down}+(U-2J)\sum_{i}n_{ix}n_{iy}+\\
J\sum_{i,\sigma,\sigma'}c_{ix\sigma}^{\dagger}c_{iy\sigma'}^{\dagger}c_{ix\sigma'}c_{iy\sigma}+J\sum_{i,\alpha\neq\beta}c_{i\alpha\up}^{\dagger}c_{i\alpha\down}^{\dagger}c_{i\beta\down}c_{i\beta\up}\label{eq:Honsite}
\end{multline}

Having derived the full interacting model, in the next sections we
discuss and classify the different instabilities of the model. By
directly decomposing the interacting term $H_{I}$ into different
irreducible channels, we obtain the effective interactions corresponding
to the possible instabilities in the particle-particle (i.e. superconducting)
and particle-hole channels in Secs. \ref{sec:sc} and \ref{sec:p-h-order},
respectively. In Sec. \ref{sec:spin-orbital}, we go one step beyond
and, in the spirit of the strong-coupling approach, include perturbatively
the kinetic Hamiltonian $H_{K}$, deriving the low-energy spin-orbital
exchange model.

\section{Superconducting instabilities and their vestigial orders \label{sec:sc}}

In this section we focus attention on the interacting Hamiltonian
$H_{I}$ of Eq. \eqref{eq:H-I} and address the question of superconductivity.
In particular, we analyze the pairing instabilities of $H_{I}$ by
decomposing the interaction into irreducible pairing channels. The
symmetry group of the normal state allows for a two-component $d$-wave
pairing channel, which gives rise to the interesting possibility of
chiral or nematic $d$-wave superconductivity. This possibility is
studied in more detail in Sec.~\ref{ssec:vestigial}.

\subsection{Decomposition of the interaction \label{ssec:decomposition}}

To decompose the interaction into irreducible pairing vertices, we
first identify the symmetry of the Cooper pairs. The full symmetry
group of the normal state, including spin rotational symmetry, is
$\mathcal{G}=D_{3}\otimes SO(3)$ (note that here we restrict to the
exact point group symmetries of TBG). This implies that the pairing
channels are labeled by the spin angular momentum $S$ of the Cooper
pair, which can take the values $S=0,1$, and the representations
$\Gamma$ of $D_{3}$, which can take the values $E\otimes E=A_{1}\oplus A_{2}\oplus E$
associated with the product of two orbitals. The decomposition of
the representation product describes the possible orbital structure
of the Cooper pair.

To proceed, we define the pair creation operator $\Pi_{i\alpha\sigma,j\beta\sigma'}^{\dagger}$
\begin{equation}
\Pi_{i\alpha\sigma,j\beta\sigma'}^{\dagger}=c_{i\alpha\sigma}^{\dagger}c_{j\beta\sigma'}^{\dagger},\label{eq:pairdef}
\end{equation}
A general pairing operator of this form can be decomposed into irreducible
pairing operators defined by the symmetry quantum numbers $(\Gamma,S,M)$.
Here $\Gamma$ denotes the point group representation and $S=0,1$
distinguishes spin-singlet and spin-triplet pairing; $M=-S,\ldots,S$.
This decomposition is given by 
\begin{equation}
\Pi_{i\alpha\sigma,j\beta\sigma'}^{\dagger}=\sum_{\Gamma}\sum_{S,M}X_{\alpha\beta}^{\Gamma}C_{\sigma\sigma'}^{SM}\Pi_{ij,\Gamma,SM}^{\dagger},\label{eq:pair-G-SM}
\end{equation}
where $C_{\sigma\sigma'}^{SM}$ are the appropriate Clebsch-Gordan
coefficients and $X_{\alpha\beta}^{\Gamma}$ are the analogues of
Clebsch-Gordan coefficients for the orbital sector. The expressions
for the latter are provided in Appendix \ref{app:decomposition}.
Note that here the sum over $\Gamma$ includes a sum over the individual
components of multi-dimensional representations, which we leave implicit
for the benefit of a more compact notation (the latter is important
and the reader is cautioned to keep this is mind).

To see how this leads to a decomposition into irreducible pairing
terms, consider the first term of $H_{I}$, Eq. \eqref{eq:H-I}, with
interaction parameters $V_{ij}^{\alpha\beta}$. Substituting Eq. \eqref{eq:pair-G-SM}
and taking sums we arrive at 
\begin{equation}
H_{I}=\frac{1}{2}\sum_{ij}\sum_{SM}\sum_{\Gamma}V_{ij}^{\Gamma}\Pi_{ij,\Gamma,SM}^{\dagger}\Pi_{ij,\Gamma,SM},\label{eq:H-V}
\end{equation}
with interaction parameters $V_{ij}^{\Gamma}$ given by 
\begin{equation}
V_{ij}^{\Gamma}=\sum_{\alpha\beta}X_{\alpha\beta}^{\Gamma}V_{ij}^{\alpha\beta}X_{\alpha\beta}^{\Gamma}.\label{eq:V-interaction}
\end{equation}
The Hamiltonian of Eq.~\eqref{eq:H-V} is diagonal in the space defined
by the spin and orbital quantum numbers $(S,M)$ and $\Gamma$. It
should be noted, however, that the interaction parameters $V_{ij}^{\Gamma}$
need not be the same for different components of the same (multi-dimensional)
representation (recall that the sum over $\Gamma$ implies a sum over
its components). This is not inconsistent with the notion of irreducible
coupling constants since these can only be defined for the full Hamiltonian
$H_{I}$. The latter includes the interaction terms $J_{1,2,3}$;
substituting the decomposition of Eq. \eqref{eq:pair-G-SM} into these
remaining terms of $H_{I}$ leads to similar expressions as Eq.~\eqref{eq:H-V}
, which can be combined to yield (details are presented in Appendix
\ref{app:decomposition}) 
\begin{equation}
H_{I}=\frac{1}{2}\sum_{ij}\sum_{SM}\sum_{\Gamma}U_{ij}^{\Gamma}\Pi_{ij,\Gamma,SM}^{\dagger}\Pi_{ij,\Gamma,SM}.\label{eq:H-U}
\end{equation}
The matrix elements $U_{ij}^{\Gamma}$ are given by the appropriate
sums of $V$ and $J_{1,2,3}$, and define the irreducible coupling
constants associated with the representation $\Gamma$.

Fermi statistics put restrictions on the allowed combinations of $\Gamma$
and $S$. This is apparent when $i=j$, in which case spin-singlet
pairing ($S=0$) can only occur for the even representations $A_{1}$
and $E$, whereas spin-triplet pairing ($S=1$) can only have $A_{2}$
symmetry. In general, the combination of $\Gamma$ and $S$ determines
whether $\Pi_{ij,\Gamma,SM}^{\dagger}$ is even or odd under the exchange
$i\leftrightarrow j$.

To illustrate the application of Eq.~\eqref{eq:H-U}, consider the
case in which the interaction terms of Eq.~\eqref{eq:H-I} are only
onsite, giving rise to Eq. \eqref{eq:Honsite}. We can express the
resulting onsite pair creation operators in the following more familiar
form:
\begin{equation}
\Pi_{\Gamma}^{\dagger}=c_{i\alpha\sigma}^{\dagger}\left[\hat{\Delta}_{\Gamma}\left(is^{y}\right)\right]_{\alpha\beta}^{\sigma\sigma'}c_{i\beta\sigma'}^{\dagger},\quad\hat{\Delta}_{\Gamma}=(\Delta_{\Gamma})_{ab}\tau^{a}s^{b},
\end{equation}
where $\hat{\Delta}_{\Gamma}$ is a matrix in orbital and spin space,
which is expanded in two sets of Pauli matrices $\tau^{a}$ and $s^{b}$
($a,b=0,x,y,z$). Here $\tau^{0}$ and $s^{0}$ are defined as the
identity. As before, $\tau^{z}=\pm1$ labels the orbital degree of
freedom and $s^{z}=\pm1$ corresponds to spin-$\up,\down$. Note that
we included explicitly the anti-symmetric tensor in spin space $(is^{y})_{\alpha\beta}=\epsilon_{\alpha\beta}$.
As mentioned, due to Fermi statistics, which can be expressed as $s^{y}\hat{\Delta}_{\Gamma}^{T}s^{y}=\hat{\Delta}_{\Gamma}$,
there are three distinct onsite pairing channels, which are uniquely
labeled by the three representations $A_{1}$, $A_{2}$, and $E$.

The onsite pair operators with $A_{1}$ and $E_{2}$ symmetry are
spin-singlet orbital-triplet states and represented by the matrices
\begin{align}
\hat{\Delta}_{A_{1}} & =1\label{eq:Delta-A1}\\
\hat{\Delta}_{E} & =(\tau^{z},\tau^{x})\label{eq:Delta-E}
\end{align}
Here the second equality expresses the fact that $E_{2}$ is two-component
representation. The pair operators with $A_{2}$ symmetry form a (orbital-singlet)
spin-triplet state transforming as $SO(3)$ under rotations in spin
space and are expressed as 
\begin{equation}
\hat{\Delta}_{A_{2}}=\tau^{y}(s^{x},s^{y},s^{z}).\label{eq:Delta-A2}
\end{equation}
Written in this form the pairing operators are not normalized. To
normalize them we multiply all matrices $\hat{\Delta}_{\Gamma}$ as
written in Eqs. \eqref{eq:Delta-A1} and \eqref{eq:Delta-A2} by a
factor $1/2\sqrt{2}$ \cite{note1}.

The coupling constants $U^{\Gamma}=U_{ii}^{\Gamma}$ of the onsite
pairing vertices, defined in Eq.~\eqref{eq:H-U}, can then be obtained
in a straightforward way. For onsite interactions, Eq. \ref{eq:Honsite},
one finds the effective interactions $U^{\Gamma}$ of the three onsite
pairings described above as (see also Table \ref{tab:symmetry_pp}):
\begin{equation}
U^{A_{1}}=U+J,\quad U^{A_{2}}=U-3J,\quad U^{E}=U-J.\label{eq_SC_vertices}
\end{equation}

\begin{table}[t]
\centering 
\begin{ruledtabular}
\begin{tabular}{lccc}
$U^{\Gamma}$  & $\Gamma=A_{1}$  & $\Gamma=A_{2}$  & $\Gamma=E_{2}$ \tabularnewline
\hline 
Singlet  & $U+J$  & $-$  & $U-J$ \tabularnewline
Triplet  & $-$  & $U-3J$  & $-$ \tabularnewline
\end{tabular}
\end{ruledtabular}

\caption{Effective interactions for the three different types of onsite particle-particle
(superconducting) orders.}
\label{tab:symmetry_pp} 
\end{table}

Note that the factor $1/2$ in \eqref{eq:H-U} was absorbed in the
normalization of the onsite pairing operators (see \cite{note1}).
Although a full analysis of the leading superconducting instabilities
is beyond the scope of this work, it is interesting to note that the
``Hund's rule'' coupling $J$ favors the $A_{2}$ and $E$ states.

To proceed with the general analysis of Eq.~\eqref{eq:H-U}, it is
convenient to go to momentum space by Fourier transforming the pair
creation operators. Specifically, we define 
\begin{equation}
\Pi_{ij}^{\dagger}=\frac{1}{N}\sum_{\bk}\Pi_{\bk\nu_{i}\nu_{j}}^{\dagger}e^{i\bk\cdot(\br_{i}-\br_{j})},\label{eq:Pi-Fourier}
\end{equation}
where $\Pi_{\bk\nu_{i}\nu_{j}}^{\dagger}=c_{\bk\nu_{i}}^{\dagger}c_{-\bk\nu_{j}}^{\dagger}$
and $\nu_{i,j}=A,B$ refers to the sublattice degree of freedom of
the honeycomb superlattice, and $N$ is the system size. In Eq.~\eqref{eq:Pi-Fourier}
spin and orbital indices have been suppressed for simplicity. Substituting
the Fourier transform into \eqref{eq:H-U}, one finds (suppressing
the spin label $S$) 
\begin{equation}
H_{I}=\frac{1}{N}\sum_{\bk\bk'}\sum_{\Gamma}\sum_{\nu\nu'}U_{\nu\nu'}^{\Gamma}(\bk'-\bk)\Pi_{\bk\nu\nu',\Gamma}^{\dagger}\Pi_{\bk'\nu\nu',\Gamma},\label{eq:H-U-k}
\end{equation}
where the momentum-dependent effective interaction $U_{\nu\nu'}^{\Gamma}(\bk)$
is given by 
\begin{equation}
U_{\nu\nu'}^{\Gamma}(\bk)=\sum_{\br_{ij}}U_{ij}^{\Gamma}e^{-i\bk\cdot\br_{ij}}.\label{eq:U(k)}
\end{equation}
This effective interaction may be compared to those of more familiar
single-band models, or of an isotropic continuum model for a Fermi
surface. Such effective interactions typically originate from (some
form of) density-density interaction. Here, apart from an additional
label $\Gamma$ associated with the orbital degree of freedom, the
effective interaction has a similar structure. In particular, as is
clear from Eq~\eqref{eq:U(k)}, it is the Fourier transform of (short-ranged)
interactions between first-, second-, and further nearest neighbor
pairs, each with their own interaction parameter.

The standard next step is to decompose $U_{\nu\nu'}^{\Gamma}(\bk'-\bk)$
into a sum over harmonics, in this case (honeycomb) lattice harmonics,
which are labeled by the symmetry quantum numbers of the lattice,
i.e., the point group representations. Such decomposition is based
on the fact that a general function $g(\bk)$ which has the symmetry
of the lattice can be expanded as $g(\bk'-\bk)=\sum_{\Gamma'}f^{\Gamma'*}(\bk')f^{\Gamma'}(\bk)$,
where $f^{\Gamma'}(\bk)$ are the lattice harmonics which transform
irreducibly.\footnote{As before, the sum over representations $\Gamma'$ includes an implicit
sum over components of multidimensional representations.} Lattice harmonics are the lattice equivalents of spherical harmonics
in isotropic systems; the latter are labeled by angular momentum quantum
numbers. An important difference with respect to isotropic systems
is the finite set of lattice symmetry quantum numbers, which implies
that distinct harmonics fall into the same channel. Once the effective
interaction \eqref{eq:U(k)} is decomposed into lattice harmonics,
the harmonics labeled by $\Gamma'$ are combined with the corresponding
pairing operators labeled by $\Gamma$ (referring to the orbitals)
to form the products $\Gamma'\otimes\Gamma$, which are reducible.
Decomposition of the product representation then yields pairing operators
fully symmetrized with respect to the symmetry group of the system.
Here we do not work this out in detail, but refer the reader to Appendix~\ref{app:U(k)}
for a more detailed discussion of decomposing \eqref{eq:U(k)}, as
well as Ref.~\onlinecite{platt2013}. Instead, we briefly showcase
the trivial case of onsite pairing in the context of Eqs. \eqref{eq:H-U-k}
and \eqref{eq:U(k)}.

The onsite component of $U_{\nu\nu'}^{\Gamma}(\bk)$ is simply given
by $U_{AA,0}^{\Gamma}=U_{BB,0}^{\Gamma}\equiv U_{0}^{\Gamma}$. What
remains to be done is to symmetrize the pairing operators with respect
to the honeycomb sublattice degree of freedom. To this end, we define
the even and odd linear combinations

\begin{align}
\Pi_{\bk AA,\Gamma}^{\dagger}+\Pi_{\bk BB,\Gamma}^{\dagger} & =\sum_{\nu\nu'}\delta_{\nu\nu'}\Pi_{\bk\nu\nu',\Gamma}^{\dagger}\\
\Pi_{\bk AA,\Gamma}^{\dagger}-\Pi_{\bk BB,\Gamma}^{\dagger} & =\sum_{\nu\nu'}\sigma_{\nu\nu'}^{z}\Pi_{\bk\nu\nu',\Gamma}^{\dagger}
\end{align}
where $\sigma^{z}=\pm1$ is an $A,B$ sublattice label. The former
is fully symmetric, whereas the latter is odd under $C_{2y}$.

At this stage it is useful to briefly connect to the recent theoretical
work on superconductivity in TBG. A number of works have addressed
the question of pairing in TBG \cite{xubalents2018,roy2018,guo2018,huang2018,ray2018,liu2018,isobe2018,peltonen2018,wumartin2018,you2018,wu2018,fidrysiak2018,kennes2018,gonzalez2018,su2018,lian2018,sherkunov2018,lin2018},
using different methods (numerical and analytical) as well as different
models. For the sake of simplicity, some authors have considered a
(two-orbital) triangular lattice model or have considered the $SU(4)$
symmetric limit of the honeycomb lattice model. Approaches have also
differed in the type of interactions included. Furthermore, while
most works focused on superconductivity from repulsive interactions,
others have explored phonon-mediated scenarios in more detail~\cite{wumartin2018,lian2018}.

Here we have presented a full symmetry-based decomposition of the
extended Hubbard interaction \eqref{eq:H-I} into pairing channels
and have obtained the corresponding coupling constants. Our starting
point is the two-orbital honeycomb lattice model, for which we do
not assume artificial higher symmetry. Notably, we make no a priori
assumptions on the range of the included interaction; the interacting
Hamiltonian \eqref{eq:H-U-k} is fully general. As a result, \eqref{eq:H-U-k}
provides the basis for studying the pairing instabilities using various
schemes. For instance, the renormalization of the interactions by
particle-hole fluctuations, treated within RPA, can be straightforwardly
included \cite{liu2018}. To this end, we derive the corresponding
decomposition in particle-hole channels in Sec.~\ref{sec:p-h-order}.

\subsection{Two-component pairing and vestigial ordering \label{ssec:vestigial}}

The existence of a two-component pairing channel, which is guaranteed
when the normal state has $D_{3}$ symmetry, merits a more detailed
discussion of the consequences of two-component superconductivity
in TBG. Since superconductors described by a two-component order parameter
break additional symmetries of the system, such as time-reversal or
rotational symmetry, they exhibit distinct signatures in experimental
probes which may be used to establish the pairing symmetry. With this
in mind we focus attention on the two-component superconducting channel
with symmetry label $E$ (hereafter denoted $E$-pairing) and consider
its phenomenology in the context of TBG. It is natural to refer to
this two-component pairing channel as $d$-wave pairing; superconductivity
with this pairing symmetry has been the focus of a number of recent
studies addressing superconductivity in TBG~\cite{xubalents2018,guo2018,huang2018,liu2018,isobe2018,fidrysiak2018,kennes2018,su2018,lin2018}

To describe an $E$-pairing state it is necessary to introduce a two-component
complex order parameter $(\eta_{1},\eta_{2})$ which transforms as
the $E$ representation of the $D_{3}$ group. The possible superconducting
ground states can be obtained by analyzing the Ginzburg-Landau expansion
of the free energy in terms of the superconducting order parameter
\cite{sigrist1991}: 
\begin{multline}
F=r(|\eta_{1}|^{2}+|\eta_{2}|^{2})+u(|\eta_{1}|^{2}+|\eta_{2}|^{2})^{2}\\
+v|\eta_{1}^{*}\eta_{2}-\eta_{2}^{*}\eta_{1}|^{2}.\label{eq:F-eta}
\end{multline}
Here $r\propto T-T_{c}$, where $T_{c}$ is the transition temperature,
and $u,v$ are fourth order expansion coefficients. The state realized
below $T_{c}$ ($r<0$) is determined by the fourth order interaction
$v$. When $v<0$, the superconducting ground state is chiral, i.e.,
time-reversal symmetry-breaking (TRSB), and given by $(\eta_{1},\eta_{2})=\eta_{0}(1,\pm i)$.
Here, $\eta_{0}$ is a complex number. A number of recent theoretical
studies have argued that this chiral $d$-wave state is favored in
TBG \cite{guo2018,huang2018,liu2018,isobe2018,fidrysiak2018,kennes2018,lin2018}.
On the other hand, when $v>0$, the superconducting ground state is
given by $(\eta_{1},\eta_{2})=\eta_{0}(\cos\phi,\sin\phi)$. Since
it preserves time-reversal symmetry but lowers the point group symmetry,
in particular threefold rotations, it is a nematic superconductor
\cite{fu2014,venderbos2016}. Importantly, the values of $\phi$ are
restricted due to the crystal symmetries. This can be seen by considering
the following sixth order term in the free energy expansion: 
\begin{equation}
F^{(6)}=\frac{\lambda}{2}\left[\left(\eta_{1}-i\eta_{2}\right)^{3}\left(\eta_{1}^{*}-i\eta_{2}^{*}\right)^{3}+\mathrm{c.c.}\right]\label{F6}
\end{equation}
For the TRSB superconducting state, this term vanishes. For the nematic
superconducting state, however, this term becomes $\lambda\left|\eta_{0}\right|^{6}\cos6\phi$,
which is minimized either by $\phi=n\pi/3$ (for $\lambda<0$) or
$\phi=\left(n+\frac{1}{2}\right)\pi/3$ (for $\lambda>0$), with integer
$n$.

The existence of a multi-component superconducting order parameter
opens the possibility of vestigial order \textemdash{} i.e. the condensation
of bilinear combinations of $\eta_{i}$ that break certain symmetries
of the lattice while preserving the $U(1)$ superconducting gauge
symmetry (for a review, see \cite{Fernandes18} and \cite{Fradkin15}).
Importantly, these bilinear combinations may condense even in the
non-superconducting state, giving rise to an ordered state that precedes
the onset of superconducting order. In the case of TBG, since it is
a two-dimensional system, superconducting phase fluctuations are very
strong and melt long-range superconducting order completely. However,
the phase with composite bilinear order is not affected by these strong
fluctuations, since it is associated with a discrete symmetry, and
thus remains as a vestige of the superconducting state.

Following Ref. \onlinecite{Fernandes18} and the analysis of the nematic
$p$-wave superconductor of Ref. \onlinecite{Schmalian18}, we identify
two possible vestigial orders, associated with the TRSB and nematic
superconducting states. In the case of a TRSB superconductor, the
composite order parameter with chiral symmetry is given by 
\begin{equation}
\psi=i\left(\eta_{1}\eta_{2}^{*}-\eta_{2}\eta_{1}^{*}\right)\equiv\boldsymbol{\eta}^{\dagger}\sigma^{y}\boldsymbol{\eta}
\end{equation}
where $\boldsymbol{\eta}=(\eta_{1},\eta_{2})^{T}$ and $\sigma^{y}$
is a Pauli matrix. It is clear that $\psi$ is a $Z_{2}$ Ising-like
order parameter, whose condensation implies TRSB (chiral order). Therefore,
the vestigial state with $\left\langle \psi\right\rangle \neq0$ but
$\left\langle \boldsymbol{\eta}\right\rangle =0$, which is expected
to take place at finite temperatures in two dimensions, is a non-superconducting
state that breaks time-reversal symmetry.

In the case of the nematic superconductor, the composite order parameter
describing nematic order has two components, which transform as partners
of the two-dimensional irreducible representation $E$: 
\begin{align}
(\Psi_{1},\Psi_{2}) & =(|\eta_{1}|^{2}-|\eta_{2}|^{2},\eta_{1}^{*}\eta_{2}+\eta_{2}^{*}\eta_{1})\nonumber \\
\boldsymbol{\Psi} & \equiv\left(\boldsymbol{\eta}^{\dagger}\sigma^{z}\boldsymbol{\eta},\,\boldsymbol{\eta}^{\dagger}\sigma^{x}\boldsymbol{\eta}\right)\label{eq:Psi-nematic}
\end{align}
Since $\boldsymbol{\Psi}$ is a composite order parameter and $\boldsymbol{\Psi}\propto(\cos2\phi,\sin2\phi)$
for $(\eta_{1},\eta_{2})=\eta_{0}(\cos\phi,\sin\phi)$, it is natural
to think of it as a $\mathbf{q}=0$ particle-hole order parameter
with $d$-wave symmetry, whose two components transform as $d_{x^{2}-y^{2}}$
and $d_{xy}$, It should be kept in mind, however, that the symmetries
of $D_{3}$ do not distinguish $p$ and $d$ waves. Importantly, the
condensation of $\boldsymbol{\Psi}$ implies that the system is no
longer invariant under an in-plane $C_{3z}$ rotation and in this
sense the ordered state can be called nematic. As a result, the vestigial
phase with $\left\langle \boldsymbol{\Psi}\right\rangle \neq0$ but
$\left\langle \boldsymbol{\eta}\right\rangle =0$ defines a nematic
phase.

At first sight, one might be tempted to identify $\boldsymbol{\Psi}$
with an XY nematic order parameter, which would not order at finite
temperatures in two dimensions due to Mermin-Wagner theorem. However,
due to crystal anisotropy $\boldsymbol{\Psi}$ is actually a $Z_{3}$
order parameter and falls in the same universality class as the 3-state
Potts model \cite{wu1982,Schmalian18}. Note that this distinguishes
it from a $Z_{2}$ Ising nematic order parameter. Indeed, writing
down the Landau free energy expansion for $\boldsymbol{\Psi}$ reveals
the existence of a cubic term: 
\begin{equation}
F_{\Psi}=r'(\Psi_{1}^{2}+\Psi_{2}^{2})+\lambda'(\Psi_{+}^{3}+\Psi_{-}^{3})+u'(\Psi_{1}^{2}+\Psi_{2}^{2})^{2},\label{eq:F-Psi}
\end{equation}
where $\Psi_{\pm}=\Psi_{1}\pm i\Psi_{2}$. Note that the existence
of a cubic term is implied by the presence of the sixth order term
\eqref{F6}; in particular, substituting \eqref{eq:Psi-nematic} into
the cubic term of \eqref{eq:F-Psi} gives \eqref{F6}.

Writing $\Psi_{+}=|\Psi|e^{i\theta}$ and expressing the cubic term
in terms of the phase $\theta$ gives $2\lambda'|\Psi|^{3}\cos3\theta$.
For $\lambda'<0$ the set of degenerate minima is given by $\theta=2n\pi/3$
with $n$ integer; for $\lambda'>0$, it is given by $\theta=(2n+1)\pi/3$.
Thus, because $\theta$ can assume three different values, $\boldsymbol{\Psi}$
is a discrete $Z_{3}$ order parameter, which can condense at finite
temperatures in two dimensions. As a result, a vestigial nematic order
is possible to be realized in TBG.  Note that the presence of the cubic order term makes the nematic transition first-order within mean-field theory \cite{Schmalian18}. However, in two dimensions, which is the case relevant for TBG, fluctuations drive the $Z_3$ transition second-order, with a small critical exponent $\beta$ for the order parameter, $\beta = 1/9$ \cite{wu1982}. The small value of $\beta$ indicates a steep onset of the nematic order parameter, which may in some experiments be similar to a jump.
Furthermore, the allowed $\theta$
values correspond to the $\pm d_{x^{2}-y^{2}}$ nematic state ($\theta=0$
and $\theta=\pi$, respectively), or to the symmetry-equivalent states
related to $\pm d_{x^{2}-y^{2}}$ by three-fold rotations. As a result,
the $d_{xy}$ nematic state ($\theta=\pm\pi/2$) is never realized,
as it is never a minimum of the free energy.

\section{Particle-hole instabilities \label{sec:p-h-order}}

In Sec.~\ref{ssec:decomposition}, for the purpose of studying superconductivity,
we decomposed the interactions into irreducible pairing (particle-particle)
channels. A similar approach can be taken to study instabilities towards
particle-hole order, such as magnetic, charge, or orbital order. Therefore,
in this section we present a decomposition of Eq. \eqref{eq:H-I}
into irreducible particle-hole channels. We begin by defining the
general particle-hole operators $\Lambda_{i\alpha\sigma,j\beta\sigma'}$
as 
\begin{equation}
\Lambda_{i\alpha\sigma,j\beta\sigma'}=c_{i\alpha\sigma}^{\dagger}c_{j\beta\sigma'},\label{eq:Lambda}
\end{equation}
which are the analogues of Eq.~\eqref{eq:pairdef}. In a manner similar
to Eq.~\eqref{eq:pair-G-SM} we decompose these operators into irreducible
particle-hole operators $\Lambda_{ij,\Gamma a}$ as 
\begin{equation}
\Lambda_{i\alpha\sigma,j\beta\sigma'}=\sum_{\Gamma}\sum_{a}Y_{\alpha\beta}^{\Gamma}\widetilde{C}_{\sigma\sigma'}^{a}\Lambda_{ij,\Gamma a},\label{eq:Lambda-G-a}
\end{equation}
where $a=0,x,y,z$ is an index for spin-singlet ($a=0$) and spin-triplet
($a=x,y,z$) particle-hole condensates. Here, the singlet and triplet
operators are defined as $\Lambda_{ij,a}=\sum_{\sigma\sigma'}c_{i\sigma}^{\dagger}s_{\sigma\sigma'}^{a}c_{j\sigma'}$,
where $s^{x,y,z}$ are the spin Pauli matrices and $s^{0}$ is the
identity. The irreducible orbital operators $\Lambda_{ij,\Gamma}$
are defined similarly; the expansion coefficients $Y_{\alpha\beta}^{\Gamma}$
and $\widetilde{C}_{\sigma\sigma'}^{a}$, which can be related to
Clebsch-Gordon coefficients, are provided in Appendix \ref{app:decomposition-ph}.
Note that the relation $\Lambda_{ij,\Gamma a}^{\dagger}=\Lambda_{ji,\Gamma a}$
holds.

\begin{table}[t]
\centering 
\begin{ruledtabular}
\begin{tabular}{lccc}
Onsite $\tilde{U}^{\Gamma a}$  & $\Gamma=A_{1}$  & $\Gamma=A_{2}$  & $\Gamma=E$ \tabularnewline
\hline 
Singlet  & $(3U-5J)/8$  & $(J-U)/8$  & $(5J-U)/8$ \tabularnewline
Triplet  & $-(U+J)/8$  & $(J-U)/8$  & $(J-U)/8$ \tabularnewline
\end{tabular}
\end{ruledtabular}
\caption{Effective interactions for onsite
particle-hole order, defined by Eq. \eqref{eq:H-U-ph}, in terms
of the interaction parameters $U$ and $J$ defined in Eq. \eqref{eq:Honsite}.
Six different channels can be distinguished based on the spin (i.e.,
singlet or triplet) and orbital structure of the particle-hole channel. }
\label{tab:symmetry_ph} 
\end{table}

Equation \eqref{eq:Lambda-G-a} is the equivalent of \eqref{eq:pair-G-SM}.
As a first step towards decomposing the interaction into particle-hole
channels, we thus proceed similarly by substituting \eqref{eq:Lambda-G-a}
into $H_{I}$. As in the case of the pairing channels, Eq. \eqref{eq:H-V},
we initially illustrate this procedure by using the density-density
terms with interaction parameters $V$. In the present case, contrary
to the pairing decomposition, we expect to obtain two terms, as there
are two ways to form particle-hole bilinears. We find for the interaction
$H_{V}$ 
\begin{multline}
H_{V}=\frac{1}{2}\sum_{ij}\sum_{\Gamma}\tilde{V}_{1,ij}^{\Gamma}\Lambda_{i,\Gamma0}\Lambda_{j,\Gamma0}\\
+\frac{1}{2}\sum_{ij}\sum_{\Gamma,a}\tilde{V}_{2,ij}^{\Gamma}\Lambda_{ij,\Gamma a}\Lambda_{ji,\Gamma a},\label{eq:H-V-ph}
\end{multline}
where the new interaction parameters $\tilde{V}_{1,ij}^{\Gamma}$
and $\tilde{V}_{2,ij}^{\Gamma}$ are given by 
\begin{equation}
\tilde{V}_{1,ij}^{\Gamma}=\sum_{\alpha\beta}Y_{\alpha\alpha}^{\Gamma}V_{ij}^{\alpha\beta}Y_{\beta\beta}^{\Gamma},\;\;\tilde{V}_{2,ij}^{\Gamma}=-\frac{1}{2}\sum_{\alpha\beta}Y_{\alpha\beta}^{\Gamma}V_{ij}^{\alpha\beta}Y_{\beta\alpha}^{\Gamma},\label{eq:V-interaction-ph}
\end{equation}
and $\Lambda_{i,\Gamma a}\equiv\Lambda_{ii,\Gamma a}=\Lambda_{i,\Gamma a}^{\dagger}$.
The first term is an interaction of pure spin-singlet onsite bilinears,
whereas the second term corresponds to the interaction of particle-hole
bilinear on bonds or sites.

The same approach applies to the exchange interaction terms $J_{1,2,3}$,
as we describe in detail in Appendix \ref{app:decomposition-ph}.
This leads to a form of $H_{I}$ given by 
\begin{multline}
H_{I}=\frac{1}{2}\sum_{ij}\sum_{\Gamma,a}\tilde{U}_{1,ij}^{\Gamma a}\Lambda_{i,\Gamma a}\Lambda_{j,\Gamma a}\\
+\frac{1}{2}\sum_{ij}\sum_{\Gamma,a}\tilde{U}_{2,ij}^{\Gamma a}\Lambda_{ij,\Gamma a}\Lambda_{ji,\Gamma a},\label{eq:H-U-ph}
\end{multline}
with effective particle-hole interactions $\tilde{U}_{1,ij}^{\Gamma a}$
and $\tilde{U}_{2,ij}^{\Gamma a}$.

Before proceeding to a more general analysis of \eqref{eq:H-U-ph},
we examine its structure in the limit where only onsite interactions
are considered, such that the interactions are parametrized by the
coefficients $U$ and $J$, see Eq.~\eqref{eq:Honsite}. As is clear
from \eqref{eq:H-U-ph}, in this case the interaction parameters can
be grouped into $\tilde{U}^{\Gamma a}=\tilde{U}_{1,ii}^{\Gamma a}+\tilde{U}_{2,ii}^{\Gamma a}$,
which then define the irreducible bare particle-hole coupling constants.
The expressions for $\tilde{U}^{\Gamma a}$ in terms of $U$ and $J$
are given in Table \ref{tab:symmetry_ph}. The particle-hole channels
corresponding to these couplings describe distinct types of particle-hole
order, in the same way that different pairing channels describe distinct
types of pairing. Spin-singlet channels may also be viewed as charge
channels, since spin-rotation invariance is preserved. For instance,
spin-singlet order with $A_{2}$ symmetry corresponds to an ordered
state with orbital magnetism, whereas singlet order with $E$ symmetry
corresponds to nematic orbital order, which breaks (three-fold) rotational
symmetry.

In a similar manner, we can explicitly express the effective
interactions $\tilde{U}_{1,ij}^{\Gamma a}$ and $\tilde{U}_{2,ij}^{\Gamma a}$
for a bond connecting a pair of distinct sites $i$
and $j$ in terms of $V$ and $J_{1,2,3}$ defined in Eq.~\eqref{eq:H-I}.
For the special case $V_{ij}^{\alpha\beta}=V_{ij}$, $J_{1,ij}^{\alpha\beta}=J_{1,ij}$,
and $J_{2}=J_{3}=0$ the result is presented in Table~\ref{tab:symmetry_ph_2}.
This particular choice of interaction parameters corresponds to the
extended Hubbard model considered in Ref.~\onlinecite{koshino2018}.

\begin{table}[t]
\centering 
\begin{ruledtabular}
\begin{tabular}{lcccc}
\multicolumn{2}{l}{Neighbors $\tilde{U}_{1,2,ij}^{\Gamma a}$} & $\Gamma=A_{1}$  & $\Gamma=A_{2}$  & $\Gamma=E$ \tabularnewline
\hline 
$\tilde{U}_{1,ij}^{\Gamma a}$  & Singlet  & $(4V_{ij}-J_{1,ij})/8$  & $-J_{1,ij}/8$  & $-J_{1,ij}/8$ \tabularnewline
 & Triplet  & $-J_{1,ij}/8$  & $-J_{1,ij}/8$  & $-J_{1,ij}/8$ \tabularnewline
\hline 
$\tilde{U}_{2,ij}^{\Gamma a}$  & Singlet  & $(4J_{1,ij}-V_{ij})/8$  & $-V_{ij}/8$  & $-V_{ij}/8$ \tabularnewline
 & Triplet  & $-V_{ij}/8$  & $-V_{ij}/8$  & $-V_{ij}/8$ \tabularnewline
\end{tabular}
\end{ruledtabular}

\caption{Effective interactions for bond particle-hole order
involving a pair of sites $(ij)$. Six different channels can be distinguished
based on the spin (i.e., singlet or triplet) and orbital structure
of the particle-hole channel. Here we have assumed $V_{ij}^{\alpha\beta}=V_{ij}$,
$J_{1,ij}^{\alpha\beta}=J_{1,ij}$, and $J_{2}=J_{3}=0$, which corresponds
to parameter values considered in Ref.~\protect\onlinecite{koshino2018}. }
\label{tab:symmetry_ph_2} 
\end{table}

We then return to a more general analysis of \eqref{eq:H-U-ph}. As
in the case of pairing, it is convenient to make use of translational
invariance and transform to momentum space. The Fourier transform
of the particle-hole operators $\Lambda_{ij,\Gamma a}$ is given by
\begin{equation}
\Lambda_{ij,\Gamma a}=\frac{1}{N}\sum_{\bq,\bk}\Lambda_{\bk\nu_{i}\nu_{j},\Gamma a}(\bq)e^{i\bq\cdot\bR_{ij}+i\bk\cdot\br_{ij}}\label{eq:Lambda-Fourier}
\end{equation}
where $\br_{ij}=\br_{i}-\br_{j}$ as before, and $\bR_{ij}=(\br_{i}+\br_{j})/2$
is the center of mass position. As in Eq. \eqref{eq:Pi-Fourier} the
Fourier transform introduces sublattice indices $\nu_{i},\nu_{j}=A,B$.
The Fourier transform of the onsite operators $\Lambda_{i,\Gamma a}$
further simplifies and is defined as $\Lambda_{\nu,\Gamma a}(\bq)=\sum_{\bk}\Lambda_{\bk\nu\nu,\Gamma a}(\bq)$.
Substituting \eqref{eq:Lambda-Fourier} into Eq.~\ref{eq:H-U-ph}
and performing the sums over site indices the interaction Hamiltonian
takes the form 
\begin{multline}
H_{I}=\frac{1}{2N}\sum_{\bq}\sum_{\Gamma,a}\tilde{U}_{1}^{\Gamma a}(\bq)\Lambda_{\Gamma a}^{\dagger}(\bq)\Lambda_{\Gamma a}(\bq)\\
+\frac{1}{2N}\sum_{\bq,\bk\bk'}\sum_{\Gamma,a}\tilde{U}_{2}^{\Gamma a}(\bk-\bk')\Lambda_{\bk',\Gamma a}^{\dagger}(\bq)\Lambda_{\bk,\Gamma a}(\bq),\label{eq:H-U-k-ph}
\end{multline}
where we have suppressed sublattice indices $\nu,\nu'$ to
avoid cumbersome expressions. The Fourier transform of the interaction
parameters $\tilde{U}_{1,ij}^{\Gamma a}$ is given by (reinstating
sublattice indices) 
\begin{equation}
\tilde{U}_{1,\nu\nu'}^{\Gamma a}(\bq)=\sum_{\br_{ij}}\tilde{U}_{1,ij}^{\Gamma a}e^{-i\bq\cdot\br_{ij}},\label{eq:U-ph(k)}
\end{equation}
and similarly for $\tilde{U}_{1,ij}^{\Gamma a}$. As may be seen from
\eqref{eq:H-U-k-ph}, the first term is now diagonal. As far as the
second term is concerned, we can follow a similar approach as in the
pairing case, see Eq.~\eqref{eq:H-U-k}, and write $\tilde{U}_{2,\nu\nu'}^{\Gamma a}(\bk-\bk')$
as a sum over lattice harmonics. The lattice harmonics are then associated
with the particle-hole operators $\Lambda_{\bk\nu\nu',\Gamma a}(\bq)$
and $\Lambda_{\bk'\nu\nu',\Gamma a}^{\dagger}(\bq)$ to form fully
symmetrized particle-hole operators.

The Hamiltonian of Eq.~\eqref{eq:H-U-k-ph} describes the effective
interactions of the particle-hole instabilities and provides a natural
framework for further analyze them. To determine which instability
is strongest within RPA, for instance, the next step is to calculate
the particle-hole bubbles in each of the irreducible
channels. This is greatly simplified by the fully symmetrized form
of the interaction.

We conclude this section by noting that an analysis of the particle-hole
instabilities in ``higher angular momentum'' channels, that is to
say, instabilities in channels corresponding to lattice harmonics
and governed by $\tilde{U}_{2}^{\Gamma a}(\bk-\bk')$, is particularly
relevant in TBG. As pointed out in Ref.~\onlinecite{thomson2018}, a
natural candidate for the ordered insulating state at quarter-filling
is a magnetic state for which the magnetic moments reside on the honeycomb
bonds. As a result, this is a bond-spin ordered state which occurs
in a particle-hole channel corresponding to nontrivial lattice harmonics.

\section{Spin-orbital exchange model at quarter filling \label{sec:spin-orbital} }

The analysis of the previous sections focused entirely
on the interacting part of the Hamiltonian $H_{I}$, classifying the
irreducible particle-particle and particle-hole channels. To obtain
a phase diagram, it is necessary to include also the kinetic term
$H_{K}$. This can be done in a controlled way in two different regimes:
weak-coupling, where $H_{I}$ is treated perturbatively, or strong-coupling,
where $H_{K}$ is treated perturbatively. The small bandwidth ($W\sim10$
meV) of the nearly flat bands in TBG does not immediately suggest
the weak-coupling approach as a natural starting point to address
electronic correlations in TBG. Indeed, estimates for the onsite Coulomb
repulsion $U$ indicate that $U\gtrsim W$ \cite{cao2018a}, placing
the system in a moderately correlated regime. To assess
this regime, in this section we opt to start from the strong-coupling
limit in which the onsite interaction $U$ is much larger then the
bandwidth.

In this case, the extended Hubbard model discussed in Sec. \ref{sec:model}
can be studied by considering the interactions first and then treating
the kinetic part as a perturbation in $\sim t/U$. This amounts to
integrating out the charge degree of freedom and results in an effective
model for the spin and orbital variables. Spin-orbital exchange models
of this Kugel-Khomskii type~\cite{kugel1973,kugel1982} have proven
rather successful in describing a large class of strongly correlated
multi-orbital systems~\cite{imada1998,dagotto2001,dagotto-book,meakawa-book}.
The key difference between the latter and TBG is the microscopic nature
of the orbital degree of freedom, which does not correspond to an
atomic orbital in TBG. Instead, the localized Wannier states of the
flat bands are associated with the Moir\'e superlattice. As a result,
the aim of this section is to explore to what extent standard approaches
from correlated multi-orbital systems can be applied to TBG.

\subsection{Derivation of the effective Hamiltonian}

To proceed, we consider the interacting Hamiltonian given by \eqref{eq:Honsite},
which only includes the onsite interactions. Restricting the interaction
to onsite terms only is an oversimplification for TBG, but necessary
for the purpose of deriving a spin-orbital model. The onsite Coulomb
repulsion of \eqref{eq:Honsite} reorganizes the Hilbert space based
on the number of electrons per site, assigning an energy cost to multiple
occupancy. Since the insulating behavior of TBG was observed for one
electron per site (or two electrons per Moir\'e supercell), we focus
on this case and define the low-energy subspace by all configurations
for which each site is singly occupied.

To obtain the effective Hamiltonian $\mathcal{H}$ we follow the standard
approach and consider virtual superexchange processes via excited
states with two electrons per site. This amounts to diagonalizing
the interacting Hamiltonian $H_{I}$ and treating the kinetic Hamiltonian
$H_{K}$ as a perturbation. In Sec.~\ref{sec:sc} we diagonalized
\eqref{eq:Honsite} in the two-particle sector and obtained the energies
of the intermediate excited states given in Table \ref{tab:symmetry_pp}.
The effective Hamiltonian can then be viewed as an expansion in $\sim t/U$.
Considering all hopping processes into the higher energy sector and
back, $\mathcal{H}$ can be expressed in the general form: 
\begin{equation}
\mathcal{H}=\mathcal{P}H_{K}^{\dagger}\frac{1}{\varepsilon_{0}-H_{I}}H_{K}\mathcal{P},\label{eq:superexchange}
\end{equation}
where $\mathcal{P}$ are projectors onto the low-energy subspace.
As is usual, the effective Hamiltonian is governed by the superexchange
energy scale $\sim t^{2}/U$. Since the virtual superexchange processes
occur on one particular bond $(ij)$, it suffices to derive the Hamiltonian
$\mathcal{H}_{ij}$ for one such bond; the full Hamiltonian $\mathcal{H}$
is given by a sum over all bonds. In principle, a superexchange coupling
of spin and orbital variables can be obtained for any pair of sites
$(ij)$ connected by $H_{K}$. In what follows, we focus attention
on the simplest case, which only includes nearest neighbor hopping.
Farther neighbor terms can be derived and analyzed analogously. In
this situation, the hopping along each bond can be parametrized by
$t_{\sigma}=t_{1}+t'_{1}$ and $t_{\pi}=t_{1}-t'_{1}$ in an appropriate
basis, see Eq. \eqref{eq:T-1} and Appendices \ref{app:wannier} and
\ref{app:symmetryD3}.

Since the microscopic Hamiltonian $H=H_{K}+H_{I}$ is $SU(2)$ spin-rotationally
invariant, the effective low-energy Hamiltonian must also be $SU(2)$
invariant, which implies that the effective Hamiltonian $\mathcal{H}_{ij}$
for a bond $(ij)$ is constructed from the projectors $\mathcal{P}_{ij}^{S=0}$
and $\mathcal{P}_{ij}^{S=1}$ onto total spin states $S=0$ and $S=1$
of the electrons connected by the bond. The projectors onto the singlet
and triplet states are given by 
\begin{equation}
\mathcal{P}_{ij}^{S=0}=\frac{1}{4}-\bS_{i}\cdot\bS_{j},\quad\mathcal{P}_{ij}^{S=1}=\frac{3}{4}+\bS_{i}\cdot\bS_{j}\label{eq:P-singlet-triplet}
\end{equation}
where $\bS_{i}$ describes the spin of site $i$.

In addition to the spin variables, the superexchange Hamiltonian acts
on the orbital variables. This action can be described by the orbital
Pauli matrices $\makebf{\tau}_{i}=(\tau_{i}^{z},\tau_{i}^{x},\tau_{i}^{y})$,
where $\tau_{i}^{z}=\pm1$ corresponds to occupancy of the $p_{x},\,p_{y}$
orbital on site $i$. Note the particular ordering of the Pauli matrices
in the definition of $\makebf{\tau}_{i}$. To capture the action of
the superexchange Hamiltonian on the orbital variables it convenient
to introduce orbital projection operators, by analogy with \eqref{eq:P-singlet-triplet}.
We introduce the projection operators $\mathcal{P}_{ij}^{\alpha\beta}$
given by 
\begin{eqnarray}
\mathcal{P}_{ij}^{xx} & = & (1+\hat{{\bf e}}_{ij}\cdot\makebf{\tau}_{i})(1+\hat{{\bf e}}_{ij}\cdot\makebf{\tau}_{j})/4,\label{eq:P-xx}\\
\mathcal{P}_{ij}^{xy} & = & (1+\hat{{\bf e}}_{ij}\cdot\makebf{\tau}_{i})(1-\hat{{\bf e}}_{ij}\cdot\makebf{\tau}_{j})/4,\label{eq:P-xy}
\end{eqnarray}
where $\hat{{\bf e}}_{ij}$ is a unit vector in the direction of the
bond $(ij)$. Therefore, $\hat{{\bf e}}_{ij}$ can take the values
$\hat{{\bf e}}_{n=1,2,3}$, which are shown in Fig.~\ref{fig:lattice_bonds}.
The projection operator $\mathcal{P}_{ij}^{xx}$, for instance, projects
on states for which the $p'_{x}=(p_{x}\hat{{\bf e}}_{x}+p_{y}\hat{{\bf e}}_{y})\cdot\hat{{\bf e}}_{ij}$
orbital is occupied on both sites $i$ and $j$.
Note that this is the $p_{x}$ orbital in a basis defined by the bond
directions $(\hat{{\bf e}}_{ij},\hat{{\bf e}}_{ij}^{\perp})$ rather
than $(\hat{{\bf e}}_{x},\hat{{\bf e}}_{y})$ \cite{wu2008}, see
Appendix \ref{app:tau} for details. In the case $\hat{{\bf e}}_{ij}=\hat{{\bf e}}_{1}=(1,0)^{T}$
the projector $\mathcal{P}_{ij}^{xx}$ takes the form $(1+\tau_{i}^{z})(1+\tau_{j}^{z})/4$.
The projector $\mathcal{P}_{ij}^{xy}$ projects on states for which
the $p'_{x}=(p_{x}\hat{{\bf e}}_{x}+p_{y}\hat{{\bf e}}_{y})\cdot\hat{{\bf e}}_{ij}$
orbital is occupied on site $i$ and the $p'_{y}=(p_{x}\hat{{\bf e}}_{x}+p_{y}\hat{{\bf e}}_{y})\cdot\hat{{\bf e}}_{ij}^{\perp}$
orbital is occupied on site $j$ (both in a bond-dependent basis).
The projection operators $\mathcal{P}_{ij}^{yy}$ and $\mathcal{P}_{ij}^{yx}$
are obtained from \eqref{eq:P-xx} and \eqref{eq:P-xy} by inverting
the signs.

To describe all superexchange processes one must also account for
the possibility that orbital flavors are flipped or exchanged. For
this purpose we introduce operators that flip the orbital occupation
of the sites $i$ and $j$; these operators are given by 
\begin{eqnarray}
\mathcal{Q}_{ij} & = & (\tau_{i}^{+}\tau_{j}^{+}+\tau_{i}^{-}\tau_{j}^{-})/2,\label{eq:Qpp}\\
\bar{\mathcal{Q}}_{ij} & = & (\tau_{i}^{+}\tau_{j}^{-}+\tau_{i}^{-}\tau_{j}^{+})/2,\label{eq:Qpm}
\end{eqnarray}
where $\tau_{i}^{\pm}$ and $\tau_{i}^{\pm}$ flip the orbital occupation
on site $i$ and $j$ in a basis defined by the bond directions $(\hat{{\bf e}}_{ij},\hat{{\bf e}}_{ij}^{\perp})$,
as before. For a bond along $\hat{{\bf e}}_{ij}=\hat{{\bf e}}_{1}$
the operator $\tau_{i}^{\pm}$ takes the form $\tau_{i}^{\pm}=\tau_{i}^{x}\pm i\tau_{i}^{y}$
(see Appendix \ref{app:tau}). Clearly, the $\mathcal{Q}_{ij}$ matrix
elements are non-zero only in the subspace of equal occupation, whereas
$\bar{\mathcal{Q}}_{ij}$ only acts within the subspace of opposite
orbital occupation.

Making use of these operators and carefully examining all superexchange
processes to obtain the correct coefficients, we find that the nearest
neighbor spin-orbital superexchange Hamiltonian $\mathcal{H}$ is
given by
\begin{widetext}
\begin{multline}
\mathcal{H}=\sum_{\langle ij\rangle}\left\{ \frac{1}{U-3J}\mathcal{P}_{ij}^{S=1}\left[t_{\sigma}t_{\pi}\bar{\mathcal{Q}}_{ij}-(t_{\sigma}^{2}+t_{\pi}^{2})(\mathcal{P}_{ij}^{xy}+\mathcal{P}_{ij}^{yx})\right]-\frac{1}{U+J}\mathcal{P}_{ij}^{S=0}\left(t_{\sigma}t_{\pi}\mathcal{Q}_{ij}+2t_{\sigma}^{2}\mathcal{P}_{ij}^{xx}+2t_{\pi}^{2}\mathcal{P}_{ij}^{yy}\right)\right.\\
\left.+\frac{1}{U-J}\mathcal{P}_{ij}^{S=0}\left[t_{\sigma}t_{\pi}(\mathcal{Q}_{ij}-\bar{\mathcal{Q}}_{ij})-2t_{\sigma}^{2}\mathcal{P}_{ij}^{xx}-2t_{\pi}^{2}\mathcal{P}_{ij}^{yy}-(t_{\sigma}^{2}+t_{\pi}^{2})(\mathcal{P}_{ij}^{xy}+\mathcal{P}_{ij}^{yx})\right]\right\} .\label{eq:HijNN}
\end{multline}
\end{widetext}

Here the sum is over honeycomb nearest neighbor sites $\langle ij\rangle$.
In its most general form given by \eqref{eq:HijNN} the Hamiltonian
describes a rather complicated coupling between spin and orbital variables,
parametrized by the two hopping integrals $t_{\sigma,\pi}$ and the
interaction terms $U,J$. This Hamiltonian can be compared to similar
spin-orbital Hamiltonians obtained in the context of correlated multi-orbital
models for transition-metal oxides \cite{feiner1997,oles2000}. 

In the present case, while \eqref{eq:HijNN} includes nearest neighbor couplings only, the superexchange Hamiltonian can be systematically extended to include farther neighbor spin-orbital superexchange couplings. This will generate superexchange terms of a similar type as in Eq. \eqref{eq:HijNN}, but for bonds $(ij)$ corresponding to second- and farther nearest neighbor sites. Using the machinery developed in our work it is in principle straightforward to obtain these additional terms by including hopping processes such as Eqs. \eqref{eq:T-2} and \eqref{eq:T-5}  in $H_K$ of \eqref{eq:superexchange}, but is expected to introduce frustration. A detailed study of the resulting phase diagram is beyond the scope of this work.

A natural first step to study \eqref{eq:HijNN} is to consider a mean-field
theory and replace the spin and orbital operators by classical variables.
This is the approach we take there.

\subsection{Mean-field solution in the isotropic limit \label{ssec:limits}}

While a full phase diagram for arbitrary values of $t_{\sigma}$ and
$t_{\pi}$ can in principle be obtained by, for instance, Monte Carlo
simulations, this is beyond the scope of our work. Rather, we develop
a mean-field theory based on an assumption which directly derives
from the reported properties of TBG. Both first-principles as well
as tight-binding calculations show that the low-energy bands of TBG
are well-described by the approximation $t_{\sigma}\approx t_{\pi}$
\cite{koshino2018}. Therefore, here we focus on the isotropic case
$t_{\sigma}=t_{\pi}\equiv t$, for which the spin-orbital Hamiltonian
\eqref{eq:HijNN} simplifies and reads as 
\begin{multline}
\mathcal{H}=\sum_{\langle ij\rangle}\left\{ \frac{t^{2}}{\left(U-3J\right)}\left(\frac{3}{4}+\bS_{i}\cdot\bS_{j}\right)\left(\makebf{\tau}_{i}\cdot\makebf{\tau}_{j}-1\right)\right.\\
-\frac{t^{2}}{U+J}\left(\frac{1}{4}-\bS_{i}\cdot\bS_{j}\right)\left(1+\makebf{\tau}_{i}\cdot\makebf{\tau}_{j}-2\tau_{i}^{y}\tau_{j}^{y}\right)\\
\left.-\frac{2t^{2}}{U-J}\left(\frac{1}{4}-\bS_{i}\cdot\bS_{j}\right)\left(\tau_{i}^{y}\tau_{j}^{y}+1\right)\right\} .\label{eq:H_isotropic}
\end{multline}
This Hamiltonian clearly reflects the higher $U(1)$
orbital symmetry that results from the neglecting the hopping anisotropy. In this form, the Hamiltonian bears resemblance to an $SU(4)$ symmetric spin-orbital model on the hyperhoneycomb lattice~\cite{natori2018}.

Before proceeding, let us briefly review the meaning of the different
degrees of freedom appearing in this Hamiltonian. A finite expectation
value $\left\langle \bS_{i}\right\rangle $ simply implies long-range
magnetic order, since $\bS_{i}$ is simply the spin at site $i$,
whose magnitude is here set to $1/2$. A finite expectation value
$\left\langle \makebf{\tau}_{i}\right\rangle $ implies some form
of orbital order, which depends on the direction of $\makebf{\tau}_{i}$
(its magnitude here is set to $1$). A finite $\left\langle \tau_{i}^{z}\right\rangle $
implies that the occupation of the $p_{x}$ and $p_{y}$ orbitals
are not the same in site $i$. This breaks rotational symmetry and
is therefore an orbital-nematic order. The same is true for $\left\langle \tau_{i}^{x}\right\rangle $,
but with the difference that $p_{x}+p_{y}$ and $p_{x}-p_{y}$ orbitals
are split in energy. Therefore, it is convenient to construct the
two-dimensional vector $\left\langle \makebf{\tau}_{i}^{\parallel}\right\rangle =\left\langle \tau_{i}^{x}\right\rangle \hat{\mathbf{x}}+\left\langle \tau_{i}^{z}\right\rangle \hat{\mathbf{z}}$,
which behaves as an XY nematic order parameter. 
In contrast
to $\left\langle \makebf{\tau}_{i}^{\parallel}\right\rangle $, a
finite $\left\langle \tau_{i}^{y}\right\rangle $ does not break rotational
symmetry but instead breaks time-reversal symmetry by selecting one
of the two orbital angular momentum eigenstates $p_{x}\pm ip_{y}$.
Consequently, a finite $\left\langle \tau_{i}^{y}\right\rangle $
implies long-range orbital-magnetic order.

Because the honeycomb superlattice is bipartite, we can find the mean-field
classical ground state by computing the classical energy of a single
bond, $E_{\mathrm{bond}}$. Since the Hamiltonian \eqref{eq:H_isotropic}
is $SU(2)$ invariant in spin-space, there are only two possible classical
spin ground states, ferromagnetic (FM) or antiferromagnetic (AFM).
We can thus find the orbital ground states in these two cases and
compare their energies to find the minimum.

Let us start with the AFM case. Defining $\Delta=t^{2}/U$, the bond
energy is given by: 
\begin{equation}
\frac{E_{\mathrm{bond}}^{(\mathrm{AFM)}}}{\Delta}=E_{0}^{(\mathrm{AFM)}}+K_{\parallel}\makebf{\tau}_{i}^{\parallel}\cdot\makebf{\tau}_{j}^{\parallel}+K_{y}\tau_{i}^{y}\tau_{j}^{y}\label{bond_AFM}
\end{equation}
where we defined: 
\begin{align}
E_{0}^{(\mathrm{AFM)}} & =\frac{2U\left(J^{2}+2JU-U^{2}\right)}{\left(U^{2}-J^{2}\right)\left(U-3J\right)}\nonumber \\
K_{\parallel} & =\frac{2JU\left(U-J\right)}{\left(U^{2}-J^{2}\right)(U-3J)}\nonumber \\
K_{y} & =\frac{4J^{2}U}{\left(U^{2}-J^{2}\right)(U-3J)}\label{aux_bond_AFM}
\end{align}

Before we proceed, we first need to discuss the range of $J/U$ values
that is reasonable. Since $U'=U-2J$, in order to have $U'>0$, we
must have $J/U<1/2$. Here, we allow $J$ to be negative as well,
which would imply violation of Hund's first rule. This was also proposed
in the context of TBG in Ref. \cite{dodaro2018}. Consequenly, in
what follows, we consider the range $-1/2<J/U<1/2$.

The orbital ground state can be obtained by analyzing the orbital
exchange constants $K_{\parallel}$ and $K_{y}$ as function of $J$.
It follows that $\left|K_{\parallel}\right|\geq\left|K_{y}\right|$
for $-1/2<J/U<1/3$. Thus, in this range, the energy is minimized
by an orbital-nematic configuration. Since $K_{\parallel}<0$ for
$J<0$, this gives ferro-orbital (FO) nematic order. On the other
hand, because $K_{\parallel}>0$ for $J>0$, we obtain antiferro-orbital
(AFO) nematic order. Similarly, because $\left|K_{\parallel}\right|<\left|K_{y}\right|$
for $1/3<J/U<1/2$, the configuration that minimizes the bond energy
is orbital-magnetic order. As $K_{y}<0$ in this range, we obtain
a ferro-orbital magnetic order.

Now let us consider the FM case. The bond energy is:

\begin{equation}
\frac{E_{\mathrm{bond}}^{(\mathrm{FM)}}}{\Delta}=E_{0}^{(\mathrm{FM)}}+K\makebf{\tau}_{i}\cdot\makebf{\tau}_{j}\label{bond_FM}
\end{equation}
with:

\begin{align}
E_{0}^{(\mathrm{FM)}} & =-\frac{U}{U-3J}\nonumber \\
K & =\frac{U}{U-3J}\label{aux_bond_FM}
\end{align}

Note that the FM bond energy is invariant under $SU(2)$ rotations
in orbital space. This ``accidental'' symmetry stems from the approximations
we employed to derive the effective Hamiltonian, and will likely be
removed if farther-neighbor hoppings are included. In any case, there
is a degeneracy in this situation between orbital-nematic and orbital-magnetic
orders. For this reason, herefater we will refer to this configuration
as $SU(2)$ orbital order.

\begin{figure}
\begin{centering}
\includegraphics[width=\columnwidth]{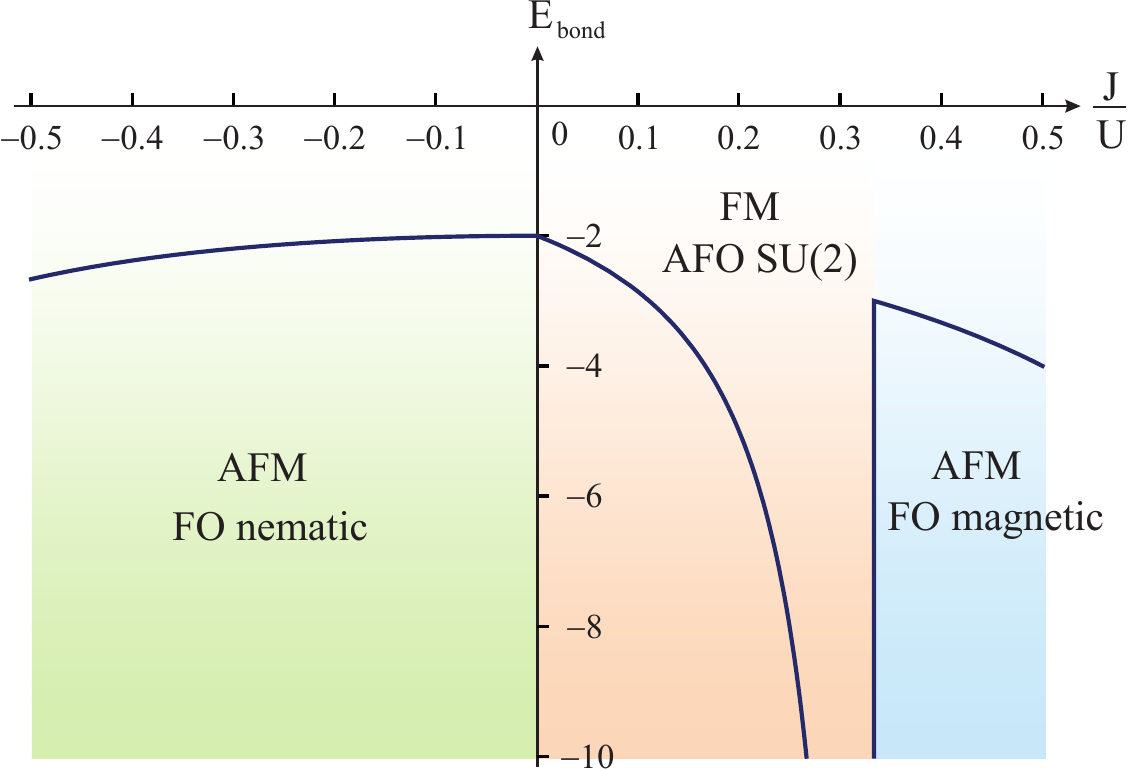} 
\par\end{centering}
\caption{Phase diagram of the classical mean-field solution of the spin-orbital
exchange model in the isotropic case ($t_{\sigma}=t_{\pi}$), obtained
by minimizing the bond energy $E_{\mathrm{bond}}$ (here plotted in
units of $\Delta=t^{2}/U$) as function of the ratio $J/U$. AFM refers
to antiferromagnetic order, FM to ferromagnetic order, FO to ferro-orbital
order, and AFO to antiferro-orbital order. For $J<0$, the orbital
order lowers the point group symmetry of the honeycomb lattice, and is thus nematic (panel $\bf{A}$ in Fig. \ref{fig:orderings}). For $0<J<U/3$, there is an enlarged $SU(2)$ symmetry in the
orbital degrees of freedom, and the orbital order can be either nematic
or magnetic (panel $\bf{B}$ in Fig. \ref{fig:orderings} illustrates the nematic case). For $J>U/3$, the system has orbital-magnetic
order (panel $\bf{C}$ in Fig. \ref{fig:orderings}). \label{fig:Phase-diagram}}
\end{figure}

Minimization of the bond energy \eqref{bond_FM} is straightforward:
for $J/U<1/3$, the orbital-exchange coefficient $K>0$ and we obtain
anti-ferro $SU(2)$ orbital order. On the other hand, for $J/U>1/3$,
we find $SU(2)$ ferro-orbital order, since $K<0$.

Having minimized the bond energies of the AFM and FM spin configurations,
we compare them to find the global bond-energy minimum. The result
is shown in Fig. \eqref{fig:Phase-diagram}, and comprises three regimes:
for $-1/2<J/U<0$, the configuration that minimizes $E_{\mathrm{bond}}$
is an antiferro-magnetic (AFM) and ferro-orbital (FO) nematic order.
For $0<J/U<1/3$, the bond energy is minimized by a ferromagnetic
(FM) and anti-ferro (AFO) $SU(2)$ orbital order. Finally, for $1/3<J/U<1/2$,
the system's configuration corresponds to AFM and ferro-orbital (FO)
magnetic order. Note that in all cases translational symmetry is broken.

We note that our strong-coupling expansion is formally not valid in
the vicinity of $J/U=1/3$, since in this case one of the denominators
of the effective Hamiltonian \eqref{eq:HijNN} diverges. Note also
that, for $J=0$, the system has additional symmetries, signaled here
by the fact that different configurations minimize the bond energy.

\section{Concluding remarks\label{sec:conclusion}}

In this paper, we analyzed the possible electronic orders arising from the two-orbital extended Hubbard model on the
honeycomb lattice, which has been proposed to describe the nearly-flat
bands of TBG. First, we presented a general framework to decompose
the several interaction terms into different irreducible particle-particle
and particle-hole channels. Although such a framework is suitable
for both weak-coupling and strong-coupling calculations, here we focused
on the latter. As a result, we derived a spin-orbital exchange model
for the quarter-filling Mott insulating state. Its mean-field solution
in the isotropic limit unveils a rich intertwinement between orbital
and spin degrees of freedom, analogous to the physics of certain correlated
multi-orbital transition metal oxides. We also discussed the possibility
of vestigial superconducting phases, which are likely to be realized
in TBG if the ground state is $d$-wave or $p$-wave, given the two-dimensional
character of TBG. While further experiments are needed to shed light
on the types of electronic order realized in TBG, the general framework
established here provides a solid starting point to assess the impact
of correlations on the spin, charge, and orbital degrees of freedom
of this system.

\begin{acknowledgments}
We would like to thank Z. Addison, L. Fu, P. Jarillo-Herrero, J. Kang,
E. J. Mele, L. Rademaker, T. Senthil, and O. Vafek for fruitful discussions. RMF
was supported by the U.S. Department of Energy, Office of Science,
Basic Energy Sciences, under Award No. DE-SC0012336. 
\end{acknowledgments}

\appendix

\section{Rotations in Wannier orbital space \label{app:wannier}}

The Wannier orbital states of the TBG honeycomb superlattice model
proposed in Refs. \onlinecite{yuan2018,kang2018,koshino2018} have
$p$-wave symmetry and transform as partners of the $E$ representation
of $D_{3}$. The operators $c_{x,y}^{\dagger}$ create electrons in
the $p_{x,y}$ Wannier states, which are defined with respect to the
$x$ and $y$ axes, i.e., a basis defined by $\hat{{\bf e}}_{x},\hat{{\bf e}}_{y}$.
We are free to choose a different basis corresponding to the rotated
vectors $\ephi,\ephi^{\perp}$ defined in \eqref{eq:e-theta}. The
rotated orbitals $p'_{x,y}$ can be expressed as $p'_{x}=(p_{x}\hat{{\bf e}}_{x}+p_{y}\hat{{\bf e}}_{y})\cdot\ephi$
and $p'_{y}=(p_{x}\hat{{\bf e}}_{x}+p_{y}\hat{{\bf e}}_{y})\cdot\ephi^{\perp}$.
This defines a rotation matrix $U_{\varphi}\equiv e^{-i\varphi\tau^{y}}$,
corresponding to a rotation by an angle $\varphi$ about the $z$
axis: 
\begin{equation}
\begin{pmatrix}p'_{x}\\
p'_{y}
\end{pmatrix}=U_{\varphi}^{\dagger}\begin{pmatrix}p_{x}\\
p_{y}
\end{pmatrix}.\label{appeq:rotation}
\end{equation}

The operators creating (annihilating) electrons in the rotated orbitals
$p'_{x,y}$ are then given by $c^{\dagger}U_{\varphi}$ ($U_{\varphi}^{\dagger}c$).
The matrix $U_{\varphi}$ is a representation of rotations $C_{\varphi z}$
about the $z$ axis generated by $\tau^{y}$. Recall that $U_{\varphi}$
is not a symmetry for general $\varphi$, but only for $\varphi_{n}=2\pi n/3$
in the case of the $D_{3}$ point group.

The rotations of the orbitals given in Eq. \eqref{appeq:rotation}
imply a rotation of the Pauli matrices $\makebf{\tau}$. Consider
first the pair of Pauli matrices $(\tau^{z},\tau^{x})$. Under rotations
in orbital space the Pauli matrices transform as 
\begin{eqnarray}
U_{\varphi}\tau^{z}U_{\varphi}^{\dagger} & = & \cos2\varphi\,\tau^{z}+\sin2\varphi\,\tau^{x},\label{app:tauz-rotation}\\
U_{\varphi}\tau^{x}U_{\varphi}^{\dagger} & = & -\sin2\varphi\,\tau^{z}+\cos2\varphi\,\tau^{x}.\label{app:taux-rotation}
\end{eqnarray}
This shows that the two Pauli matrices transform as partners under
rotations and that they have $d$-wave symmetry: 
\begin{equation}
C_{\varphi z}\;:\;\begin{pmatrix}\tau^{z}\\
\tau^{x}
\end{pmatrix}\to U_{2\varphi}^{\dagger}\begin{pmatrix}\tau^{z}\\
\tau^{x}
\end{pmatrix}.
\end{equation}
We can also define the matrices $U_{x}=\tau_{z}$ and $U_{y}=-\tau_{z}$
that represent the two-fold rotations about the $x$ axis ($C_{2x}$)
and $y$ axis ($C_{2y}$), respectively. Under either of these transformations,
$(\tau^{z},\tau^{x})\to(\tau^{z},-\tau^{x})$. Meanwhile, the Pauli
matrix $\tau^{y}$ is invariant under $C_{\varphi z}$ rotations but
odd under $C_{2y}$ and $C_{2x}$ rotations. This implies that $(\tau^{z},\tau^{x})$
have $E$ symmetry under $D_{3}$ and $\tau^{y}$ has $A_{2}$ symmetry.

The form of the rotation matrix $U_{\varphi}\equiv e^{-i\varphi\tau^{y}}$
implies that it is diagonal in a basis in which $\tau^{y}$ is diagonal.
This basis is defined by the orbitals complex orbitals $p_{\pm}=p_{x}\pm ip_{y}$,
which are eigenvectors of the angular momentum projections $L_{z}=\pm1$.
If we define $c_{\pm}$ as the operators corresponding to $p_{\pm}$,
then one has $C_{\varphi z}\;:\;c_{\pm}\to e^{\pm i\varphi}c_{\pm}$.
This implies that if the terms in the kinetic Hamiltonian, Eq. \eqref{eq:H-K},
do not couple $c_{+}$ and $c_{-}$, which is only true for a specific
set of (fine-tuned) hopping parameters, the kinetic Hamiltonian $H_{K}$
has a larger $U(1)$ symmetry given by $C_{\varphi z}$. Since the
orbitals $p_{\pm}$ can be related to the valley degrees of freedom
of the constituent graphene layers \cite{koshino2018} this larger
symmetry can be associated with a $U(1)$ valley symmetry.

\section{Hopping matrix symmetry constraints for $D_{3}$ model \label{app:symmetryD3}}

In this Appendix, we review the symmetry constraints on the hopping
matrices discussed in Ref.~\onlinecite{kang2018} using a different
formalism.

The kinetic Hamiltonian of Eq. \eqref{eq:H-K} defines the hopping
matrices $\hat{T}(\br_{ij})$, where $\br_{ij}$ is the distance between
sites forming the bond $(ij)$. It is natural to group the set of
hopping matrices into subsets defined by fixed distance $\br_{ij}$,
which is a grouping based on nearest neighbors, and we introduce the
index $\gamma$ to denote the $\gamma$-th nearest neighbor bonds.
That is, $\gamma=1,2,3$ denotes the first, second, and third nearest
neighbors. We then rewrite the set of hopping matrices as $\hat{T}_{n}^{(\gamma)}$,
where $n=1,\dots,N_{\gamma}$ is an index for all the $\gamma$-th
nearest neighbors, of which there are $N_{\gamma}$.

For given $\gamma$ one may then obtain symmetry constraints for $\hat{T}_{n}^{(\gamma)}$,
from which the number of independent hopping parameters can be determined.
As an example, consider the first-nearest neighbor ($\gamma=1$) hopping
matrix for $n=1$. Due to time-reversal symmetry there exists a gauge
in which all matrix elements of $\hat{T}_{1}^{(1)}$ are real and
the hopping matrix can be expanded in orbital Pauli matrices as 
\begin{equation}
\hat{T}_{1}^{(1)}=t_{10}+t_{1z}\tau^{z}+t_{1x}\tau^{x}+it_{1y}\tau^{y},\label{appeq:T1-expand}
\end{equation}
where $(t_{10},t_{1x},t_{1y},t_{1z})$ are four real parameters. The
two-fold rotation $C_{2y}$ gives rise to constraints on these parameters.
Abbreviating $\hat{T}_{n=1}^{(1)}$ as $\hat{T}$ for simplicity,
the constraints can be stated as 
\begin{equation}
C_{2y}\;\to\;\tau^{z}\hat{T}\tau^{z}=\hat{T}^{\dagger}.
\label{appeq:T1-C2y}
\end{equation}
The appearance of $\hat{T}^{\dagger}$ on the right hand side of the
constraint \eqref{appeq:T1-C2y} is due to the fact that $C_{2y}$
exchanges the sites connected by the bond. The constraint \eqref{appeq:T1-C2y}
forces $t_{1x}=0$, which would lead to three independent hopping
parameters. As noted in Ref. \onlinecite{kang2018}, however, with
a redefinition of the basis of the two Wannier states one of these
can be absorbed. It is natural to choose $t_{1y}$ and this leads
to Eq. \eqref{eq:T-1} with $(t_{10},t_{1z})\equiv(t_{1},t'_{1})$.
Since we have now fixed the basis of the Wannier states no further
symmetry-allowed hopping parameters (of further neighbor bonds) can
be absorbed by redefinition.

The two remaining first-nearest neighbor hopping matrices $T_{2,3}^{(1)}$
follow directly from $T_{1}^{(1)}$ by performing three-fold rotations:
\begin{equation}
\hat{T}_{2}^{(1)}=U_{\varphi_{2}}\hat{T}_{1}^{(1)}U_{\varphi_{2}}^{\dagger},\quad\hat{T}_{3}^{(1)}=U_{\varphi_{3}}\hat{T}_{1}^{(1)}U_{\varphi_{3}}^{\dagger},\label{appeq:T1-rotate}
\end{equation}
where $\varphi_{n}=2\pi(n-1)/3$ are the angles of the nearest neighbor
unit vectors (see Sec. \ref{sec:model}).

The same analysis can be applied to any of the other inter-sublattice
hoppings matrices, i.e., those matrices corresponding to bonds connecting
sites on different sublattices. We take the third-nearest neighbor
hopping (i.e., across a hexagon) as an example and expand 
\begin{equation}
\hat{T}_{1}^{(3)}=t_{30}+t_{3z}\tau^{z}+t_{3x}\tau^{x}+it_{3y}\tau^{y},\label{appeq:T3-expand}
\end{equation}
where $(t_{30},t_{3x},t_{3y},t_{3z})$ are again four real parameters.
Now abbreviating $\hat{T}_{n=1}^{(3)}$ as $\hat{T}$ we find the
constraint from $C_{2y}$ as 
\begin{equation}
C_{2y}\;\to\;\tau^{z}\hat{T}\tau^{z}=\hat{T}^{\dagger}.\label{appeq:T3-C2y}
\end{equation}
This is the same constraint as \eqref{appeq:T1-C2y} and we conclude
that $t_{3x}=0$. As a result, $\hat{T}_{1}^{(3)}$ has three real
parameters and is given by $\hat{T}_{1}^{(3)}=t_{3}+t'_{3}\tau^{z}+it''_{3}\tau^{y}$.
The remaining third-nearest neighbor hopping matrices are found by
rotation as in Eq.~\eqref{appeq:T1-rotate}.

Next, consider intra-sublattice hoppings associated with bonds connecting
sites on the same sublattice. The simplest example is second-nearest
neighbor ($\gamma=2$) hopping. (This is first-nearest neighbor hopping
on the triangular sublattice.) Again, we start from $n=1$, i.e.,
$\hat{T}_{1}^{(2)}$, which corresponds to the second-nearest neighbor
bond along the direction of $\ba_{1}$ in Fig. \ref{fig:lattice_bonds}.
As before we expand 
\begin{equation}
\hat{T}_{1}^{(2)}=t_{20}+t_{2z}\tau^{z}+t_{2x}\tau^{x}+it_{2y}\tau^{y},\label{appeq:T2-expand}
\end{equation}
with real coefficients. To determine the symmetry constraints on the
coefficients we must account for the two sublattices $A$ and $B$.
We abbreviate $\hat{T}_{1}^{(2)}$ on the $A$ ($B$) sublattice as
$\hat{T}_{A}$ ($\hat{T}_{B}$) and find that the constraints from
the twofold rotation $C_{2y}$ is given by 
\begin{equation}
C_{2y}\;\to\;\tau^{z}\hat{T}_{A}\tau^{z}=\hat{T}_{B}.\label{appeq:T2-C2y}
\end{equation}
This equation does not give rise to constraints on the hopping parameters
on one sublattice, but instead relates the hopping parameters on the
two sublattices. In particular, $(t_{20},t_{2z})$ are identical on
the two sublattices, whereas $(t_{2x},t_{2y})$ have opposite sign.

As a second example of intra-sublattice hopping, consider fifth-nearest
neighbor hopping. Fifth-nearest neighbor hopping, which is second-nearest
neighbor on the triangular sublattices, has played an important role
in previous work \cite{yuan2018,koshino2018}. In particular, it was
identified as being responsible for the splitting of bands along $\Gamma$\textendash $M$
in models with an additional $U(1)$ symmetry. Consider the bond defined
by the lattice vector $\ba_{3}-\ba_{2}$; we define the corresponding
hopping matrix $\hat{T}_{1}^{(5)}$ and expand it as before as 
\begin{equation}
\hat{T}_{1}^{(5)}=t_{50}+t_{5z}\tau^{z}+t_{5x}\tau^{x}+it_{5y}\tau^{y},\label{appeq:T5-expand}
\end{equation}
For simplicity, we once more abbreviate $\hat{T}_{1}^{(5)}$ on the
$A$ ($B$) sublattice as $\hat{T}_{A}$ ($\hat{T}_{B}$). The constraints
from the twofold rotation $C_{2y}$ now reads as 
\begin{equation}
C_{2y}\;\to\;\tau^{z}\hat{T}_{A}\tau^{z}=\hat{T}_{B}^{\dagger}.\label{appeq:T5-C2y}
\end{equation}
Comparison with Eq.~\eqref{appeq:T2-C2y} shows that~\eqref{appeq:T5-C2y}
leads to a different relation between $t_{5y}$ on the two sublattices.
Specifically, one finds that $(t_{50},t_{5z},t_{5y})$ are identical
on the two sublattices, whereas only $t_{5x}$ has opposite sign.
It is precisely this property of $t_{5y}$ which is responsible for
the splitting of bands along $\Gamma$\textendash $M$.

\section{Decomposition into irreducible pairing channels \label{app:decomposition}}

The pair creation operator $\Pi_{i\alpha\sigma,j\beta\sigma'}^{\dagger}$
is defined as 
\begin{equation}
\Pi_{i\alpha\sigma,j\beta\sigma'}^{\dagger}=c_{i\alpha\sigma}^{\dagger}c_{j\beta\sigma'}^{\dagger},\label{appeq:Pi-pair}
\end{equation}
such that a two-particle state $\ket{i\alpha\sigma;j\beta\sigma'}$
is given by $\ket{i\alpha\sigma;j\beta\sigma'}=\Pi_{i\alpha\sigma,j\beta\sigma'}^{\dagger}\ket{0}$.
Note that this definition implies $\Pi_{i\alpha\sigma,j\beta\sigma'}=c_{j\beta\sigma'}c_{i\alpha\sigma}$.
A general pairing operator can be decomposed into irreducible pairing
operators with symmetry quantum numbers $(\Gamma,S,M)$ as 
\begin{equation}
\Pi_{i\alpha\sigma,j\beta\sigma'}^{\dagger}=\sum_{\Gamma}\sum_{S,M}X_{\alpha\beta}^{\Gamma}C_{\sigma\sigma'}^{SM}\Pi_{ij,\Gamma,SM}^{\dagger},\label{appeq:pair-Gam-SM}
\end{equation}
where $C_{\sigma\sigma'}^{SM}=C_{\frac{1}{2}\sigma\frac{1}{2}\sigma'}^{SM}=\bracket{\frac{1}{2}\frac{1}{2};SM}{\frac{1}{2}\sigma;\frac{1}{2}\sigma'}$
are Clebsch-Gordan coefficients. Here $S=0$ corresponds to spin-singlet
pairing and $S=1$ corresponds to spin-triplet pairing, in which case
$M$ takes values $M=-1,0,1$.

Similar to singlet and triplet pairing operators, the operators $\Pi_{ij,\Gamma}^{\dagger}$
(suppressing spin for simplicity) are symmetrized in orbital space
and are thus labeled by point group representations $\Gamma\in\{A_{1},A_{2},E\}$.
Note that $E$ is a two-dimensional $d$-wave channel. The irreducible
pairing operators $\Pi_{ij,\Gamma}^{\dagger}$ are given by 
\begin{gather}
\Pi_{ij,A_{1}}^{\dagger}=\sum_{\alpha\beta}\frac{\delta_{\alpha\beta}}{\sqrt{2}}c_{i\alpha}^{\dagger}c_{j\beta}^{\dagger},\;\;\Pi_{ij,A_{2}}^{\dagger}=\sum_{\alpha\beta}\frac{\epsilon_{\alpha\beta}}{\sqrt{2}}c_{i\alpha}^{\dagger}c_{j\beta}^{\dagger},\\
(\Pi_{ij,E_{1}}^{\dagger},\Pi_{ij,E_{2}}^{\dagger})=\frac{1}{\sqrt{2}}\sum_{\alpha\beta}(\tau_{\alpha\beta}^{z},\tau_{\alpha\beta}^{x})c_{i\alpha}^{\dagger}c_{j\beta}^{\dagger}.\label{appeq:orbital-pair}
\end{gather}
The coefficients $X_{\alpha\beta}^{\Gamma}$ in Eq. \eqref{appeq:pair-Gam-SM}
are the analogs of Clebsch-Gordon coefficients for the orbital sector;
they are given by 
\begin{gather}
X_{\alpha\beta}^{A_{1}}=\frac{1}{\sqrt{2}}\delta_{\alpha\beta},\quad X_{\alpha\beta}^{A_{2}}=\frac{1}{\sqrt{2}}\epsilon_{\alpha\beta},\\
(X_{\alpha\beta}^{E_{1}},X_{\alpha\beta}^{E_{2}})=\frac{1}{\sqrt{2}}(\tau_{\alpha\beta}^{z},\tau_{\alpha\beta}^{x}).\label{eq:pairSM}
\end{gather}

Fermi statistics imposes constraints on the decomposition of Eq. \eqref{appeq:pair-Gam-SM},
in particular on the set of quantum numbers $(\Gamma,S,M)$. Spin-singlet
and spin-triplet states are anti-symmetric and symmetric with respect
to particle exchange, respectively; similarly, states with $A_{2}$
symmetry are anti-symmetric and states with $A_{1}$ or $E$ symmetry
are symmetric. As a result, when $i=j$ spin-singlet states can only
have $A_{1}$ or $E$ symmetry, whereas spin-triplet states must have
$A_{2}$ symmetry. In general, one has the relation 
\begin{equation}
\Pi_{ij,\Gamma,SM}^{\dagger}=(-1)^{p_{\Gamma}+p_{S}+1}\Pi_{ji,\Gamma,SM}^{\dagger},\label{eq:Fermi}
\end{equation}
where $p_{S}$ is the parity of the spin state (i.e., $p_{0}=1$ and
$p_{1}=0$) and $p_{\Gamma}$ is the parity of the orbital state (i.e.,
$p_{A_{2}}=1$ and zero otherwise).


Substituting Eq. \eqref{appeq:pair-Gam-SM} into $H_{I}$ of Eq. \eqref{eq:H-I}
we arrive at the form 
\begin{equation}
H_{I}=\sum_{ij}\sum_{SM}\sum_{\Gamma}U_{ij}^{\Gamma}\Pi_{ij,\Gamma,SM}^{\dagger}\Pi_{ij,\Gamma',SM},\label{appeq:H-Uij}
\end{equation}
where matrix elements $U_{ij}^{\Gamma}$ are defined as 
\begin{equation}
U_{ij}^{\Gamma}=V_{ij}^{\Gamma}+J_{1,ij}^{\Gamma}+J_{2,ij}^{\Gamma}+J_{3,ij}^{\Gamma}.\label{appeq:U-GG}
\end{equation}
The matrix elements $V_{ij}^{\Gamma}$ are given by Eq. \eqref{eq:V-interaction};
the expressions for the remaining matrix elements are 
\begin{eqnarray}
J_{1,ij}^{\Gamma} & = & (-1)^{p_{\Gamma}+p_{S}+1}\sum_{\alpha\beta}X_{\alpha\beta}^{\Gamma}(J_{1})_{ij}^{\alpha\beta}X_{\alpha\beta}^{\Gamma},\label{eq:J1-GG}\\
J_{2,ij}^{\Gamma} & = & (-1)^{p_{\Gamma}+p_{S}+1}\sum_{\alpha\beta}X_{\alpha\beta}^{\Gamma}(J_{2})_{ij}^{\alpha\beta}X_{\beta\alpha}^{\Gamma},\label{eq:J2-GG}\\
J_{3,ij}^{\Gamma} & = & (-1)^{p_{\Gamma}+p_{S}+1}\sum_{\alpha\beta}X_{\alpha\alpha}^{\Gamma}(J_{3})_{ij}^{\alpha\beta}X_{\beta\beta}^{\Gamma}.\label{eq:J3-GG}
\end{eqnarray}
At this point, it is important to recall that the sum over $\Gamma$
in Eq.~\eqref{appeq:H-Uij} {[}and, obviously, in Eq.~\eqref{eq:H-U}{]}
includes an implicit sum over the components of multidimensional representations;
in the present case only $E$ is multidimensional. The irreducible
coupling constants $U_{ij}^{\Gamma}$ given by Eq.~\eqref{appeq:U-GG}
are a property of the pairing channel and therefore of the representation.
As a result, they must be the same for all components of a representation
and are appropriately labeled by $\Gamma$. Importantly, however,
each of the interaction parameters on the right hand side of \eqref{appeq:U-GG}
need \emph{not} be the same for all components of a representation,
only their sum. In particular, the expressions of Eqs.~\eqref{eq:J1-GG}\textendash \eqref{eq:J3-GG}
should be evaluated for each component of a representation $\Gamma$.
This fact is obscured by adopting a more compact notation, but the
reader is cautioned to keep this in mind.

The requirement that $U_{ij}^{\Gamma}$ defines the coupling constant
of a representation $\Gamma$ gives rise to a constraint on the interaction
parameters $V$ and $J_{1,2,3}$, since their sum must be proportional
to the identity within each representation. The consequences of such
constraint are exemplified by the onsite Hamiltonian of Eq. \eqref{eq:Honsite},
which is specified in terms of only two interaction energy scales.

\section{Further decomposition of Eq.~\eqref{eq:U(k)} \label{app:U(k)}}

The decomposition of $U_{\nu\nu'}^{\Gamma}(\bk'-\bk)$ follows the
standard scheme for identifying the irreducible pairing channels in
a system with symmetry group $\mathcal{G}$. As explained in Sec.~\ref{sec:sc},
the vertex function $U_{\nu\nu'}^{\Gamma}(\bk)$ is the Fourier transform
of the interactions between pairs, which in practice will be short-ranged
and thus limited to the first few nearest neighbors. Using the notation
of Appendix \ref{app:symmetryD3}, the interaction parameters can
be denoted $U_{\gamma}^{\Gamma}$, where $\gamma=1,2,3$ corresponds
to first, second, and third nearest neighbors; $U_{0}^{\Gamma}$ defines
the onsite interactions. As an example, the term in $U_{\nu\nu'}^{\Gamma}(\bk)$
corresponding to first-nearest neighbor interactions takes the form
\begin{equation}
U_{AB,1}^{\Gamma}(\bk)=U_{BA,1}^{\Gamma*}(\bk)=U_{1}^{\Gamma}\sum_{n}\exp(i\bk\cdot\bd_{n}),\label{appeq:U_1(k)}
\end{equation}
where $\bd_{n=1,2,3}$ denote the nearest neighbor vectors in the
direction $\hat{{\bf e}}_{n}$, see Fig.~\ref{fig:lattice_bonds}.
Similarly, the second-nearest neighbor interactions are given by 
\begin{equation}
U_{AA,2}^{\Gamma}(\bk)=U_{BB,2}^{\Gamma}(\bk)=U_{2}^{\Gamma}\sum_{n}\cos\bk\cdot\ba_{n},\label{appeq:U_2(k)}
\end{equation}
where $\ba_{n=1,2,3}$ are the three primitive lattice vectors shown
in Fig.~\ref{fig:lattice_bonds}.

For each $\gamma$, the next step is to decompose $U_{\nu\nu'}^{\Gamma}(\bk)$
into lattice harmonics $f^{\Gamma'}(\bk)$ as 
\begin{equation}
U_{\gamma}^{\Gamma}(\bk'-\bk)=U_{\gamma}^{\Gamma}\sum_{\Gamma'}f^{\Gamma'*}(\bk')f^{\Gamma'}(\bk),\label{appeq:U(k-k')}
\end{equation}
where we have suppressed the sublattice $\nu\nu'$ for simplicity.
The sum over $\Gamma'$ should be understood as a sum over all distinct
symmetry quantum numbers, which in particular includes a sum over
the components of multidimensional representations. To showcase \eqref{appeq:U(k-k')},
consider the second nearest neighbor interactions given by \eqref{appeq:U_2(k)}.
In this case $U_{2}^{\Gamma}(\bk'-\bk)$ is decomposed into a sum
over six lattice harmonics given by 
\begin{eqnarray}
f^{A_{1},+}(\bk) & = & \sum_{n}\cos\bk\cdot\ba_{n},\label{appeq:f-nnn-1}\\
f^{E_{1},+}(\bk) & = & \text{Re}\sum_{n}e^{i4\pi(n-1)/3}\cos\bk\cdot\ba_{n},\label{appeq:f-nnn-2}\\
f^{E_{2},+}(\bk) & = & \text{Im}\sum_{n}e^{i4\pi(n-1)/3}\cos\bk\cdot\ba_{n},\label{appeq:f-nnn-3}
\end{eqnarray}
as well as $f^{A_{1},-}(\bk)$ and $f^{E,-}(\bk)$ obtained from \eqref{appeq:f-nnn-1}\textendash \eqref{appeq:f-nnn-3}
by replacing $\cos\bk\cdot\ba_{n}$ with $\sin\bk\cdot\ba_{n}$. Note
that the functions $f^{\pm}(\bk)$ have the property $f^{\pm}(-\bk)=\pm f^{\pm}(\bk)$.
The parity under $\bk\to-\bk$ is important, since Fermi statistics
implies 
\begin{equation}
\Pi_{\bk\Gamma,SM}^{\dagger}=(-1)^{p_{\Gamma}+p_{S}+1}\Pi_{-\bk\Gamma,SM}^{\dagger}.\label{appeq:Fermi-2}
\end{equation}

The final step is to form irreducible momentum space pairing operators
by coupling the lattice harmonics to the orbital degree of freedom.
This amounts to taking the product $\Gamma'\otimes\Gamma$, where
the first refers to the lattice and second to the orbital degree of
freedom, and decomposing it into irreducible terms. This exactly analogous
to forming total angular pairing operators in spin-orbit coupled systems,
in which spin is locked to the lattice. Here, instead, the orbital
degree of freedom is (intrinsically) locked to the lattice.


\section{Decomposition into irreducible particle-hole channels \label{app:decomposition-ph}}

The particle-hole pair operators $\Lambda_{i\alpha\sigma,j\beta\sigma'}$
are defined in \eqref{eq:Lambda} and their decomposition in terms
of orbital and spin symmetrized pair operators is given by Eq.~\eqref{eq:Lambda-G-a}.
The spin-singlet/triplet and the coefficients $C_{\sigma\sigma'}^{a}$
are given by (suppressing orbital indices) 
\begin{equation}
\Lambda_{ij,a}=\sum_{\sigma\sigma'}c_{i\sigma}^{\dagger}s_{\sigma\sigma'}^{a}c_{j\sigma'},\quad\widetilde{C}_{\sigma\sigma'}^{a}=\frac{1}{2}s_{\sigma'\sigma}^{a}.\label{appeq:Lambda-spin}
\end{equation}
Here $s^{x,y,z}$ are a set of Pauli matrices acting on the electron
spin and $s^{0}$ is the identity; recall that $a=0,x,y,z$. The symmetrized
orbital operators are defined as (suppressing spin indices) 
\begin{gather}
\Lambda_{ij,A_{1}}=\sum_{\alpha\beta}\delta_{\alpha\beta}c_{i\alpha}^{\dagger}c_{j\beta},\quad\Lambda_{ij,A_{2}}=\sum_{\alpha\beta}c_{i\alpha}^{\dagger}\tau_{\alpha\beta}^{y}c_{j\beta},\\
(\Lambda_{ij,E_{1}},\Lambda_{ij,E_{2}})=\sum_{\alpha\beta}(\tau_{\alpha\beta}^{z},\tau_{\alpha\beta}^{x})c_{i\alpha}^{\dagger}c_{j\beta},\label{appeq:Lambda-orbital}
\end{gather}
and the orbital expansion coefficients $Y_{\alpha\beta}^{\Gamma}$
are given by 
\begin{gather}
Y_{\alpha\beta}^{A_{1}}=\frac{1}{2}\delta_{\alpha\beta},\quad Y_{\alpha\beta}^{A_{2}}=\frac{1}{2}\tau_{\beta\alpha}^{y},\\
(Y_{\alpha\beta}^{E_{1}},Y_{\alpha\beta}^{E_{2}})=\frac{1}{2}(\tau_{\alpha\beta}^{z},\tau_{\alpha\beta}^{x}).\label{eq:pairSM}
\end{gather}
With these definitions one has $\Lambda_{ji,\Gamma a}=\Lambda_{ij,\Gamma a}^{\dagger}$,
which implies that 
\begin{equation}
\Lambda_{ji,\Gamma a}\Lambda_{ij,\Gamma a}=\Lambda_{ij,\Gamma a}^{\dagger}\Lambda_{ij,\Gamma a}=\left|\Lambda_{ij,\Gamma a}\right|^{2}.\label{appeq:Lambda-dagger}
\end{equation}

Using the expansions coefficients and Eq.~\eqref{eq:Lambda-G-a}
the interaction parameters $\tilde{U}_{1,ij}^{\Gamma a}$ and $\tilde{U}_{2,ij}^{\Gamma a}$
of Eq.~\eqref{eq:H-U-ph} can be determined. In contrast to the pairing
case, here the interaction parameters depend on the spin structure
of the symmetrized particle-hole operators. We must distinguish singlet
interactions ($a=0$) and singlet interactions ($a=x,y,z$). For the
case $\tilde{U}_{1,ij}^{\Gamma a}$ we find 
\begin{eqnarray}
\tilde{U}_{1,ij}^{\Gamma,0} & = & \tilde{V}_{1,ij}^{\Gamma}+\tilde{J}_{11,ij}^{\Gamma}+\tilde{J}_{21,ij}^{\Gamma}+\tilde{J}_{31,ij}^{\Gamma},\label{appeq:U1-S=00003D0}\\
\tilde{U}_{1,ij}^{\Gamma x,y,z} & = & \tilde{J}_{11,ij}^{\Gamma}+\tilde{J}_{21,ij}^{\Gamma}+\tilde{J}_{31,ij}^{\Gamma},\label{appeq:U1-S=00003D1}
\end{eqnarray}
whereas for the parameters $\tilde{U}_{2,ij}^{\Gamma a}$ we find
\begin{eqnarray}
\tilde{U}_{2,ij}^{\Gamma0} & = & \tilde{V}_{2,ij}^{\Gamma}+\tilde{J}_{12,ij}^{\Gamma}+\tilde{J}_{22,ij}^{\Gamma}+\tilde{J}_{32,ij}^{\Gamma},\label{appeq:U2-S=00003D0}\\
\tilde{U}_{2,ij}^{\Gamma x,y,z} & = & \tilde{V}_{2,ij}^{\Gamma}.\label{appeq:U2-S=00003D1}
\end{eqnarray}
The parameters on the right hand side are given by 
\begin{eqnarray}
\tilde{J}_{11,ij}^{\Gamma} & = & -\frac{1}{2}\sum_{\alpha\beta}Y_{\alpha\beta}^{\Gamma}(J_{1})_{ij}^{\alpha\beta}Y_{\beta\alpha}^{\Gamma},\\
\tilde{J}_{12,ij}^{\Gamma} & = & \sum_{\alpha\beta}Y_{\alpha\alpha}^{\Gamma}(J_{1})_{ij}^{\alpha\beta}Y_{\beta\beta}^{\Gamma},\label{eq:J1-interaction-ph}
\end{eqnarray}
for the $J_{1}$ exchange interaction, 
\begin{eqnarray}
\tilde{J}_{21,ij}^{\Gamma} & = & -\frac{1}{2}\sum_{\alpha\beta}Y_{\alpha\alpha}^{\Gamma}(J_{2})_{ij}^{\alpha\beta}Y_{\beta\beta}^{\Gamma},\\
\tilde{J}_{22,ij}^{\Gamma} & = & \sum_{\alpha\beta}Y_{\alpha\beta}^{\Gamma}(J_{2})_{ij}^{\alpha\beta}Y_{\beta\alpha}^{\Gamma},\label{eq:J2-interaction-ph}
\end{eqnarray}
for the $J_{2}$ exchange interaction, and 
\begin{eqnarray}
\tilde{J}_{31,ij}^{\Gamma} & = & -\frac{1}{2}\sum_{\alpha\beta}Y_{\alpha\beta}^{\Gamma}(J_{3})_{ij}^{\alpha\beta}Y_{\beta\alpha}^{\Gamma},\\
\tilde{J}_{32,ij}^{\Gamma} & = & \sum_{\alpha\beta}Y_{\alpha\beta}^{\Gamma}(J_{3})_{ij}^{\alpha\beta}Y_{\alpha\beta}^{\Gamma},\label{eq:J3-interaction-ph}
\end{eqnarray}
for the $J_{3}$ exchange interaction.

\section{Orbital $\tau$ variables in the chiral basis \label{app:tau}}

It is convenient to rearrange the orbital Pauli matrices $\makebf{\tau}_{i}=(\tau_{i}^{x},\tau_{i}^{y},\tau_{i}^{z})$
in a way which exploits their transformation properties under rotations
in orbital space (see also Appendix \ref{app:wannier}). To make this
explicit we can relabel the Pauli matrices as 
\begin{equation}
\makebf{\tau}_{i}\to(\tau_{i}^{1},\tau_{i}^{2},\tau_{i}^{3})\equiv(\tau_{i}^{z},\tau_{i}^{x},\tau_{i}^{y}).\label{appeq:tau-basis}
\end{equation}
In this way $\tau^{3}$ generates rotations about the $z$ axis and
$(\tau_{i}^{1},\tau_{i}^{2})$ transform as a nematic director under
such rotations. To make see this clearly, recall Eqs.~\eqref{app:tauz-rotation}
and \eqref{app:taux-rotation}, which show how $\makebf{\tau}_{i}$
transforms under rotations of the orbitals. In terms of the redefined
$\makebf{\tau}_{i}$ variables of \eqref{appeq:tau-basis} the rotation
of $\makebf{\tau}_{i}$ can be expressed on the simple form 
\begin{equation}
U_{\varphi}\tau_{i}^{1}U_{\varphi}^{\dagger}=\hat{{\bf e}}_{2\varphi}\cdot\makebf{\tau}_{i},\quad U_{\varphi}\tau^{2}U_{\varphi}^{\dagger}=\hat{{\bf e}}_{2\varphi}^{\perp}\cdot\makebf{\tau}_{i},
\end{equation}
where the use of the dot product now has a natural interpretation.
Since the orbitals $p_{ix,y}$ are eigenstates of $\tau_{i}^{1}$,
the rotated orbitals $p'_{ix,y}$ of Eq. \eqref{appeq:rotation} are
eigenstates of $\hat{{\bf e}}_{2\varphi}\cdot\makebf{\tau}_{i}$.

As mentioned, the redefinition of \eqref{appeq:tau-basis} is designed
so that $\tau^{3}$ generates rotations about the $z$ axis. Rotations
by $\pi$ about the $x$ axis are represented by $\tau^{1}$ and rotations
by $\pi$ about the bisector of the $x$ and $y$ axes are represented
by $\tau^{2}$. This implies that under rotations by $\pi$ about
the $x$ axis the $\tau$ variables change as $\tau^{1}\to\tau^{1},\tau^{2,3}\to-\tau^{2,3}$.
Therefore, if we rotate the orbitals by $180^{\circ}$ about the $x$
axis, which changes $(p_{x},p_{y})$ to $(p_{x},-p_{y})$, the Pauli
matrices $\tau^{1}$ and $\tau^{2}$ transform under rotations by
$\varphi$ as: $\tau^{1}\to\hat{{\bf e}}_{-2\varphi}\cdot\makebf{\tau}$
and $\tau^{2}\to\hat{{\bf e}}_{-2\varphi}^{\perp}\cdot\makebf{\tau}$
\cite{wu2008}. This is very useful since $\varphi=-2\varphi$ for
$\varphi=0,2\pi/3,4\pi/3$, which are precisely the angles corresponding
to the three nearest neighbor bond directions $\hat{{\bf e}}_{n=1,2,3}$
of the honeycomb lattice (see Fig. \ref{fig:lattice_bonds}). As a
result, the eigenstates of $\hat{{\bf e}}_{n}\cdot\makebf{\tau}$
are precisely the $p'_{x}$ and $-p'_{y}$ orbitals along bond $\hat{{\bf e}}_{n}$.

With the relabeling of $\makebf{\tau}_{i}$ matrices and the basis
transformation of the orbitals it is then a simple matter to construct
the orbital projection operators of Eqs. \eqref{eq:P-xx} and \eqref{eq:P-xy}.
Note first that 
\begin{equation}
\mathcal{P}_{i}^{x,y}=\frac{1}{2}(1\pm\hat{{\bf e}}_{ij}\cdot\makebf{\tau}_{i}),
\end{equation}
are projection operators which project onto the orbitals $p'_{ix}=(p_{ix}\hat{{\bf e}}_{x}+p_{iy}\hat{{\bf e}}_{y})\cdot\hat{{\bf e}}_{ij}$
and $p'_{iy}=(p_{ix}\hat{{\bf e}}_{x}+p_{iy}\hat{{\bf e}}_{y})\cdot\hat{{\bf e}}_{ij}^{\perp}$.
The same is true for site $j$: $\mathcal{P}_{j}^{x,y}=\frac{1}{2}(1\pm\hat{{\bf e}}_{ij}\cdot\makebf{\tau}_{j})$.
From these we define the four projection operators $\mathcal{P}_{ij}^{xx}$,
$\mathcal{P}_{ij}^{yy}$, $\mathcal{P}_{ij}^{xy}$, and $\mathcal{P}_{ij}^{yx}$
given by 
\begin{equation}
\mathcal{P}_{ij}^{x,y;x,y}=\frac{1}{4}(1\pm\hat{{\bf e}}_{ij}\cdot\makebf{\tau}_{i})(1\pm\hat{{\bf e}}_{ij}\cdot\makebf{\tau}_{j}).
\end{equation}

The orbital flip operators of Eqs.~\eqref{eq:Qpp} and \eqref{eq:Qpm}
are defined based on the same conventions. In particular, for two
nearest neighbor sites $i$ and $j$ the orbital raising and lowering
operators are defined as 
\begin{equation}
\tau_{i}^{\pm}={\bf e}_{ij}^{\perp}\cdot\makebf{\tau}_{i}\pm i\tau_{i}^{3},\quad\tau_{j}^{\pm}={\bf e}_{ij}^{\perp}\cdot\makebf{\tau}_{j}\pm i\tau_{j}^{3}.
\end{equation}
For the case ${\bf e}_{ij}={\bf e}_{n=1}$ this reduces to $\tau_{i}^{\pm}=\tau_{i}^{2}\pm i\tau_{i}^{3}=\tau_{i}^{x}\pm i\tau_{i}^{y}$.




\begin{thebibliography}{10}
\bibitem{note3} This correspondence is only exact when starting from
a low-energy continuum, which is the approach taken in Ref. \onlinecite{koshino2018}.
When the Wannier states are derived from a microscopic tight-binding
model for commensurate TBG, as in Ref.~\onlinecite{kang2018}, there
is no exact valley symmetry.
\bibitem{note1} Due to the fact that these are onsite pair operators,
the normalization differs by a factor $1/\sqrt{2}$ from the definition
of the general irreducible pair operators in Eq. \eqref{eq:pair-G-SM}.
\end{thebibliography}
\end{document}